\documentclass[preprint,12pt]{elsarticle}

\usepackage{amssymb}
\usepackage{hyperref}
\hypersetup{
    hidelinks
}
\usepackage{floatrow}
\usepackage{algorithm}
\usepackage{graphicx}
\usepackage{algpseudocode}
\usepackage{amsmath}
\usepackage[size=footnotesize]{subcaption}
\usepackage[textsize=tiny,disable]{todonotes}
\graphicspath{{figures/}}
\usepackage{cleveref}
\crefname{appendix}{}{}
\usepackage{soul}
\usepackage{geometry}
\usepackage{algorithmicx}
\usepackage{booktabs}
\usepackage{longtable}
\usepackage{siunitx}
\usepackage{stmaryrd}
\usepackage{algpseudocode}
\usepackage{algorithm}
\usepackage{adjustbox}
\usepackage{color, colortbl}
\usepackage{multirow}
\usepackage{xcolor}
\usepackage{ulem}
\usepackage{pgfgantt}
\usepackage{microtype}

\newcommand{\x}{\mathbf{x}}
\newcommand{\n}{\mathbf{n}}

\journal{}

\newcommand\notes[1]{\textcolor{red}{#1}} 
\renewcommand\notes[1]{}

\let\tempone\itemize
\let\temptwo\enditemize
\renewenvironment{itemize}{\tempone\addtolength{\itemsep}{-0.5\baselineskip}}{\temptwo}

\graphicspath{{./Figures}}

\usepackage{siunitx}  

\usepackage{listings}

\newcommand*\subtxt[1]{_{\mathrm{#1}}} 
    \DeclareRobustCommand\_{\ifmmode\expandafter\subtxt\else\textunderscore\fi}

\newcommand*\subsubtxt[1]{_{_{\mathrm{#1}}}}
    \DeclareRobustCommand\-{\ifmmode\expandafter\subsubtxt\else\textunderscore\fi}

\begin{document}

\begin{frontmatter}



\title{Stabilizing the unstructured Volume-of-Fluid method for capillary flows in microstructures using artificial viscosity}


\author[inst1]{Luise Nagel}
\affiliation[inst1]{organization={Robert Bosch GmbH, Corporate Research},
            addressline={Robert-Bosch-Campus 1}, 
            city={Renningen},
            postcode={71272}, 
            country={Germany}}

\author[inst1]{Anja Lippert\corref{corr}}
\ead{Anja.Lippert@bosch.com}
\cortext[corr]{Corresponding author. }

\author[inst1]{Tobias Tolle}

\author[inst1]{Ronny Leonhardt}

\author[inst2]{Huijie Zhang}
\affiliation[inst2]{organization={Robert Bosch GmbH, Automotive Electronics},
            addressline={Markwiesenstrasse 46}, 
            city={Reutlingen},
            postcode={72770}, 
            country={Germany}}

\author[inst3]{Tomislav Mari\'c}
\affiliation[inst3]{organization={Technical University of Darmstadt, Mathematical Modeling and Analysis},
            addressline={Alarich-Weiss-Straße 10}, 
            city={Darmstadt},
            postcode={64287}, 
            country={Germany}}

\begin{abstract}
\color{blue}
\textit{This is the author manuscript of an article \cite{nagel_stabilizing_2024} published in Experimental and Computational Multiphase Flow, available at \href{https://www.sciopen.com/article/10.1007/s42757-023-0181-y}{sciopen.com/article/10.1007/s42757-023-0181-y}. 
Please cite the published journal article \cite{nagel_stabilizing_2024} when referring to this manuscript.} 
\\
\color{black}
Parasitic currents still pose a significant challenge for the investigation of two-phase flow in Lab-on-Chip (LoC) applications with Volume-of-Fluid (VoF) simulations.
To counter the impact of such spurious velocity fields in the vicinity of the fluid interface, this work presents an implementation of an artificial interface viscosity model in OpenFOAM.
The model is introduced as an additional dampening term in the momentum conservation equation.
It is implemented as a fvOption, allowing for its simple application to existing VoF solvers.
Validation is performed with hydrodynamic and wetting cases, in which constant artificial viscosity values are prescribed to examine the sensitivity of the solution to the artificial dampening.
The artificial viscosity model shows promising results in reducing spurious currents for two considered geometrical VoF solvers, namely interIsoFoam and InterFlow.
It is found that the influence of the artificial viscosity heavily depends on the fluid properties.
Applying the model to simulations of an interface traversing through microcavities relevant in LoC applications, experimental results of the interface progression are predicted well, while spurious currents are effectively reduced by approximately one order of magnitude due to the artificial viscosity model.
The code is publicly available on GitHub\footnote{\href{https://github.com/boschresearch/sepMultiphaseFoam/tree/publications/ArtificialInterfaceViscosity}{https://github.com/boschresearch/sepMultiphaseFoam/tree/publications/ArtificialInterfaceViscosity}}.
\end{abstract}




\end{frontmatter}


\section{Introduction}
\label{sec:introduction}
\notes{
Disclaimer to Anja and Tomislav: Everything displayed in gray are my notes. Remove them from the PDF by uncommenting line 5 in style.tex.  
\\
possible journal: https://www.sciencedirect.com/journal/international-journal-of-multiphase-flow ; impact factor 4.044 (2022). If more focus on AVI modelling: https://www.sciencedirect.com/journal/computers-and-fluids ; impact factor 3.077 (2022)
}
\notes{
Introduction contents:
\begin{itemize}
    \item LoC and multiphase flow - CFD challenges with surface tension force modelling
    \item sparse available validation data for for realistic 3D flow in LoC cavities - short literature review
    \item Spurious currents and their mitigation in literature
    \item time step constraints ? here or later?
    \item short artificial viscosity literature review, reasons for using artificial viscosity
    \item mention wisps, their effect and wisp correction (cite Tomislavs thesis)
    \item core achievements of this paper
    \item Section overview
\end{itemize}
}
As a major field in healthcare research, Lab-on-Chip (LoC) technology promises rapid and automated sample analysis through miniaturization. 
LoC devices exploit complex microfluidic processes with multiple immiscible phases, species transport, and biochemical reactions to perform a wide range of bio-chemical assays directly at the point of care rather than in central laboratories.
In the sub-millimeter geometries present in LoC applications, surface tension driven two-phase flows play an important role, with one relevant example being the filling of microcavity arrays with sample liquids. 
The compartmentalization of liquids into discrete isolated nanoliter volumes allows, among others, for the application of multiplex Polymerase Chain Reaction (PCR) in molecular diagnostics~\cite{podbiel_fusing_2020}
and for separation and analysis of circulating tumor cells in blood samples~\cite{yin_microfluidics-based_2019}.
The success and efficiency of these complex processes depend heavily on the understanding of the flow during the filling process, as recently discussed in~\cite{podbiel_analytical_2020}.
%
\par 
Experimental flow investigations in this regime are classically evaluated optically and are thus limited to a 2D perspective.
To support experimental data of two-phase flows with 3D simulation insights, the Volume-of-Fluid (VoF) model within the unstructured Finite Volume Method (FVM) is widely used\notes{citation?} (cf.~\cite{maric2020unstructured} for a recent review).
Recent extensive benchmark studies on the performance of different VoF solvers were published in~\cite{lippert_benchmark_2022} for hydrodynamic cases and in~\cite{grunding_comparative_2020} and~\cite{asghar2023numerical} for wetting cases.
However, few studies exist considering the application of VoF to more complex LoC geometries.
In~\cite{padmanabhan_enhanced_2020}, 2D simulations of microcavity filling are performed to study the influence of geometry on the filling behavior, focusing mostly on stationary results. 
To the knowledge of the authors, simulations of transient 3D separated two-phase flows on microcavity structures are not present in literature.
On the experimental side, many investigations of microcavity filling focus on the overall filling behavior of arrays consisting of several hundred cavities, as investigated in~\cite{liu_rapid_2009}~and~\cite{cui_facile_2021}.
Wetting effects such as the contact line pinning at microcavity edges are only investigated for very specific geometries, as in~\cite{sposito_staggered_2017}~and~\cite{lin_scalable_2020},
leaving a need for reproducible experimental data of 3D cavity filling for validation of simulations.
%
\par The low number of reports on transient 3D simulation of surface tension dominated flows in LoC geometries can partly be attributed to spurious currents reducing the solution quality below a usable level.
Spurious currents, also known as parasitic currents, are well known to appear in the vicinity of the fluid interface in VoF simulations. 
They originate in numerical inaccuracies in the approximation of the surface tension force, more specifically in the estimation of the interface curvature.
In regimes where surface tension is dominant, spurious currents can exceed the physical velocity by up to several orders of magnitude.
While their influence on the interface movement in the simulation is often neglected for practical reasons, they can impair the quantitative investigation of velocity fields in the vicinity of the interface and inhibit the investigation of species transport in this region.
%
\par

Mitigation of spurious currents in the VoF method is an active research field.
One method to counter the effects of spurious currents is the introduction of an artificial viscosity.
This method originally stems from the goal of circumventing the capillary time step constraint, which is strict in surface tension dominated regimes~\cite{brackbill_continuum_1992}.
Since the capillary time step is associated with the explicit implementation of the surface tension force in the momentum equation, Brackbill~\cite{brackbill_continuum_1992} suggests formulating the surface tension force term implicitly.
A first implementation of a semi-implicit surface tension in the VoF formulation is proposed by Raessi et al.~\cite{raessi_semi-implicit_2009}.
They extend the model of Bänsch~\cite{bansch_finite_2001} and Hysing~\cite{hysing_new_2006}, formulated for the finite element method, to the FVM.
The resulting extension of the surface tension force takes the form of a dampening term active at the fluid interface, which is now known as an artificial viscosity model.
Denner et al.~\cite{denner_artificial_2017} extend this model and provide a thorough validation study using an in-house VoF solver~\cite{denner_fully-coupled_2014}.
By applying the artificial viscosity model, they are able to overcome the capillary time step restriction in several dynamic cases and achieve a reduction of spurious currents in the simulation of falling liquid films.
The semi-implicit model of Hysing~\cite{hysing_new_2006} is extended to the hybrid Level Set / Front Tracking method in~\cite{tolle_saample:_2020}.
\par 
This work presents an implementation of an artificial viscosity model similar to the one shown in~\cite{denner_artificial_2017} for the geometrical VoF method~\cite{scheufler_accurate_2019} in OpenFOAM~\cite{OpenFOAMcode}.
Hydrodynamic and wetting benchmarks based on recently published studies in~\cite{lippert_benchmark_2022}, \cite{denner_artificial_2017}, and \cite{asghar2023numerical} are considered for validation.
The focus is on the reduction of spurious currents, and it is shown that depending on the case setup and fluid properties, spurious currents can be reduced in magnitude and in area of influence with the artificial viscosity model.
Finally, the model is applied to a simulation of a fluid interface traversing through a microcavity array.
In comparison to own experiments, the simulation is able to predict the filling behavior of the interface in two scenarios.
It is shown that the application of the artificial viscosity model reduces spurious currents by approximately one order of magnitude while having only minimal influence on the interface movement prediction.
\par
Besides the artificial viscosity model, the removal of wisps is considered as a further aspect of this work.
Wisps are small artificial interface elements in the bulk phase which are known to reduce the solution accuracy.
In this work, a simplified version of the wisp correction proposed by Mari\'c~\cite{maric_enhanced_2018} is implemented.
In the context of the microcavity array simulations, wisps appear during the filling of the cavities.
The results show that the wisp removal in combination with the artificial viscosity model can further reduce spurious currents.
\par
The code is publicly available on GitHub\footnote{https://github.com/boschresearch/sepMultiphaseFoam/tree/publications/ArtificialInterfaceViscosity}.


\section{Mathematical and numerical modeling}
\label{sec:mathematicalMdeling}
\subsection{Mathematical modeling: The one-field formulation for two-phase flows}
\label{sec:volume-of-fluid}
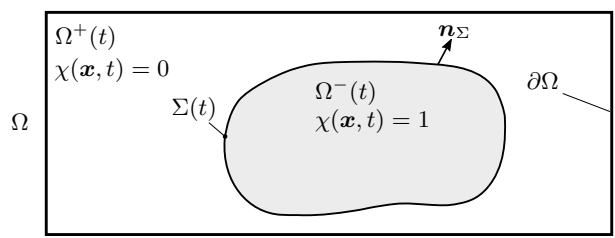
\begin{figure}[!htb]
    \centering
    \fontsize{9pt}{11pt}\selectfont
    \def\svgwidth{0.5\textwidth}
\begingroup%
  \makeatletter%
  \providecommand\color[2][]{%
    \errmessage{(Inkscape) Color is used for the text in Inkscape, but the package 'color.sty' is not loaded}%
    \renewcommand\color[2][]{}%
  }%
  \providecommand\transparent[1]{%
    \errmessage{(Inkscape) Transparency is used (non-zero) for the text in Inkscape, but the package 'transparent.sty' is not loaded}%
    \renewcommand\transparent[1]{}%
  }%
  \providecommand\rotatebox[2]{#2}%
  \newcommand*\fsize{\dimexpr\f@size pt\relax}%
  \newcommand*\lineheight[1]{\fontsize{\fsize}{#1\fsize}\selectfont}%
  \ifx\svgwidth\undefined%
    \setlength{\unitlength}{476.65983906bp}%
    \ifx\svgscale\undefined%
      \relax%
    \else%
      \setlength{\unitlength}{\unitlength * \real{\svgscale}}%
    \fi%
  \else%
    \setlength{\unitlength}{\svgwidth}%
  \fi%
  \global\let\svgwidth\undefined%
  \global\let\svgscale\undefined%
  \makeatother%
  \begin{picture}(1,0.39629091)%
    \lineheight{1}%
    \setlength\tabcolsep{0pt}%
    \put(0,0){\includegraphics[width=\unitlength,page=1]{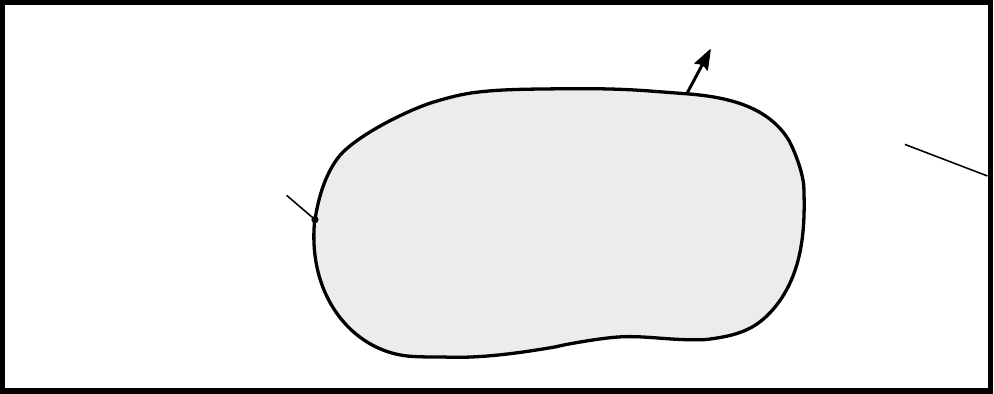}}%
    \put(0.47546785,0.24149135){\color[rgb]{0,0,0}\makebox(0,0)[lt]{\smash{\begin{tabular}[t]{l}$\Omega^-(t)$\end{tabular}}}}%
    \put(0.47546785,0.1921162){\color[rgb]{0,0,0}\makebox(0,0)[lt]{\smash{\begin{tabular}[t]{l}$\chi(\boldsymbol{x}, t) = 1$\end{tabular}}}}%
    \put(0.01721016,0.28246681){\color[rgb]{0,0,0}\makebox(0,0)[lt]{\smash{\begin{tabular}[t]{l}$\chi(\boldsymbol{x}, t) = 0$\end{tabular}}}}%
    \put(0.84833234,0.25964645){\color[rgb]{0,0,0}\makebox(0,0)[lt]{\smash{\begin{tabular}[t]{l}$\partial \Omega$\end{tabular}}}}%
    \put(0.69217011,0.35271978){\color[rgb]{0,0,0}\makebox(0,0)[lt]{\smash{\begin{tabular}[t]{l}$\boldsymbol{n}_{\Sigma}$\end{tabular}}}}%
    \put(0.01795905,0.33811759){\color[rgb]{0,0,0}\makebox(0,0)[lt]{\smash{\begin{tabular}[t]{l}$\Omega^+(t)$\end{tabular}}}}%
    \put(0.22278415,0.21292546){\color[rgb]{0,0,0}\makebox(0,0)[lt]{\smash{\begin{tabular}[t]{l}$\Sigma(t)$\end{tabular}}}}%
    \put(-0.05904447,0.18633432){\color[rgb]{0,0,0}\makebox(0,0)[lt]{\smash{\begin{tabular}[t]{l}$\Omega$\end{tabular}}}}%
  \end{picture}%
\endgroup%
\notag
    \caption{Sketch of a two-phase flows domain.}
    \label{fig:drop-gas}
\end{figure}
The geometrical Volume-of-Fluid method is based on the one-field formulation of the two-phase Navier-Stokes Equations (NSE).
To represent the two fluid phases in a domain, a phase indicator $\chi$ is defined as
\begin{align}
\chi\left(\mathbf{x}, t\right)=
    \begin{cases}
        1 & \text{in $\Omega^-(t)$}, \\
        0 & \text{in $\Omega^+(t)$}\,.
    \end{cases}
    \label{eq:phaseind}
\end{align}
Using the phase indicator, the single-field density and viscosity are modeled as
\begin{equation}
 \begin{aligned}
    \rho(\x, t)  = \chi(\x, t) \rho^- + (1 - \chi(\x, t)) \rho^+, \\
    \mu(\x, t)  = \chi(\x, t) \mu^- + (1 - \chi(\x, t)) \mu^+,
 \end{aligned}
 \label{eq:dens-visc}
\end{equation}
where $\rho^{\pm}$ and $\mu^{\pm}$ are the density and dynamic viscosity of phases $\Omega^{\pm}$, respectively.
Using~\cref{eq:dens-visc} and assuming incompressibility of both phases, the one-field NSE are given as the mass conservation equation
\begin{equation}
    \nabla\cdot \mathbf{u} = 0, 
    \label{eq:volconserv}
\end{equation}
with $\mathbf{u}$ being the velocity, and the momentum conservation equation
\begin{equation}
    \frac{\partial}{\partial t} (\rho \mathbf{u}) + \nabla\cdot(\rho \mathbf{u} \mathbf{u}) - \nabla\cdot(\mu \nabla \mathbf{u}) = -\nabla p' - \nabla \rho (\mathbf{g} \cdot \mathbf{x}) + \nabla\cdot(\mu (\nabla \mathbf{u})^T) + \mathbf{f}_\Sigma , 
    \label{eq:momentum}
\end{equation}
where $p' = p - \rho (\mathbf{g} \cdot \mathbf{x})$ is the dynamic pressure and $\mathbf{g}$ is the gravitational acceleration.
More details on the pressure formulation are given in~\cite{popinet_numerical_2018}.
The last term in~\cref{eq:momentum},
\begin{equation}
    \mathbf{f}_\Sigma:=\sigma \kappa_\Sigma \n_\Sigma \delta_\Sigma
    \label{eq:fsigma}
\end{equation}
is the surface tension force.
It consists of the constant surface tension coefficient $\sigma$, the curvature of the fluid interface $\kappa_\Sigma$, the interface normal vector $\n_\Sigma$ and the interface Dirac's distribution $\delta_\Sigma$. 
More details on the mathematical model are available in \citep{Tryggvason2011} and~\cite{popinet_numerical_2018}.

%
\subsection{Numerical modeling: The geometrical VoF method}
The geometrical Volume-of-Fluid (VoF) method based on the unstructured Finite Volume Method (FVM) is used in this work.
A detailed review of the unstructured geometrical VoF method is available in~\cite{maric2020unstructured}.
In the present work, two geometrical VoF solvers are employed, namely interIsoFoam and interFlow.
The former is a state-of-the-art solver in the OpenFOAM framework introduced in~\cite{roenby_computational_2016}.
 The latter is part of the OpenFOAM-based framework TwoPhaseFlow presented in \cite{scheufler2021twophaseflow} and \cite{TwoPhaseFlowCode}.
The two solvers are based on the IsoAdvector-plicRDF method described in~\cite{roenby_computational_2016} and~\cite{scheufler_accurate_2019} and differ in the model for the interface curvature that appears in the discretized surface tension force
\begin{align}
\mathbf{f}_{\Sigma,f} = \sigma\kappa_{\Sigma,f}(\nabla\alpha)_f.
\label{eq:csf}
\end{align}
\Cref{eq:csf} is the discretization of~\cref{eq:fsigma} based on the forced-balanced Continuum Surface Force (CSF) model which is discussed in detail in~\cite{popinet_numerical_2018}.
The surface tension force is evaluated at the face centroids of the computational cells, denoted by the index~$f$.
In~\cref{eq:csf}, $\alpha$ denotes the volume fraction which is the volume averaged phase indicator.
\par In interIsoFoam, the curvature model \notes{is based on the work of Brackbill et al.~\cite{brackbill_continuum_1992} and }reads~\cite{scheufler2021twophaseflow}
\begin{align}
    \kappa_{\Sigma, f} =-\left[\nabla\cdot\left(\frac{\nabla \alpha}{|\nabla \alpha|}\right)\right]_f\,.
    \label{eq:csf-curvature}
\end{align}
The TwoPhaseFlow framework offers two alternative models for unstructured meshes explained in full detail in~\cite{scheufler2021twophaseflow}.
While the \textit{Reconstructed Distance Function (RDF)} curvature model is based on the gradient of a geometrically constructed signed distance function, 
the \textit{fitParaboloid} model locally fits a paraboloid to an interface cell and its neighbor cells as basis for the curvature estimation.
In this work, the \textit{fitParaboloid} model is used due to its superior performance in wetting cases found in~\cite{asghar2023numerical}.
For simplicity, this solver configuration is referred to as interFlow-fPa in the following sections.
\subsection{Wisp removal}
\label{subsec:wisp}
In this work, a wisp is defined as a nearly empty or nearly full cell completely surrounded by empty, respectively full cells based on a wisp tolerance~$\epsilon\_w$. 
A conservative wisp removal algorithm for parallel computations is proposed by Mari\'c in \cite{maric_enhanced_2018}.
It consists of a process-internal wisp detection step, a process-boundary correction and a redistribution step. 
In the present work, only the first and second step are considered, meaning that identified wisps are simply removed by emptying or filling a cell, accepting the resulting small mass errors.
Further details on the wisp correction can be found in~\cite{maric_enhanced_2018}.
%
%

\color{black}{}
\section{Artificial viscosity model for separated two-phase flows}
\label{section:aviModel}
\subsection{Mathematical modeling}
\label{subsec:aviModel_mathematicalModelling}
The aim of formulating the surface tension force term~(\cref{eq:csf}) implicitly is to circumvent the capillary time step constraint, as indicated by Brackbill~\cite{brackbill_continuum_1992}.
The capillary time step describes the temporal resolution necessary to represent the shortest spatially resolved capillary waves.
It reads
\begin{equation} 
    \Delta t\_c < \sqrt{\dfrac{(\rho^- + \rho^+) (\Delta x)^{3} }{4 \pi \sigma} }.
    \label{eq:dt_capillary}
\end{equation}
Under the assumption that an implicit CSF implementation can migitate this strict critierion, the explicit surface tension force term is extended by an additional implicit term as shown in \cite{raessi_semi-implicit_2009}. 
As a consequence, in the VoF formulation, the surface tension term \cref{eq:fsigma} is extended to\notes{improve sentence.} 
\begin{equation}
    \mathbf{f}_{\Sigma} = \sigma \kappa_{\Sigma} \mathbf{n}_\Sigma \delta_\Sigma +  \mathbf{f}_{\Sigma \mathrm{I}},
    \label{eq:extended_fsigma}
\end{equation}
with the implicit term
\begin{equation} 
    \mathbf{f}\_{\Sigma I} =  \sigma \Delta t \Delta_\Sigma (\mathbf{u}) \delta\_\Sigma. 
    \label{eq:implicit_st} 
\end{equation}
Here, $\Delta t$ is the time step and $\Delta_\Sigma()$ denotes the Laplace-Beltrami operator of the interface $\Sigma$. 
\Cref{eq:implicit_st} acts as an additional shear stress in the momentum conservation \cref{eq:momentum}.
Popinet~\cite{popinet_numerical_2018} and Denner~\cite{denner_artificial_2017} note that~\cref{eq:implicit_st} can be reformulated to
\begin{equation}
    \mathbf{f}\_{\Sigma I} =  \eta_\Sigma \Delta_\Sigma (\mathbf{u})
\end{equation}
when introducing the artificial viscosity at the interface $\eta_\Sigma$ which has the dimension of a dynamic viscosity. 
Now, $\eta_\Sigma$ can be considered as a free parameter that is classically modeled as $\eta_\Sigma := \sigma \Delta t \delta_\Sigma$ as given in~\cref{eq:implicit_st}, but can also be represented by a constant parameter in relation to the physical viscosities in the domain.
Thus, the implicit surface tension expression in \cref{eq:implicit_st} can be viewed as an additional diffusion term that can be calibrated to the problem at hand.
\notes{
\begin{itemize}
    \item Capillary time step restriction can be very strict for ... $\rightarrow$ in introduction!
    \item choose a good formulation - use face-centered formulation as in csf equation, or Tobis equation from saample paper, or both?
\end{itemize}
}
\subsection{Numerical modeling and implementation}
\label{subsec:aviModel_implementation}
An implementation of the artificial viscosity model in OpenFOAM is given by Tolle et al. for the unstructured Level Set / Front Tracking method \cite{tolle_saample:_2020}.
With the implicit Euler method,~\cref{eq:extended_fsigma} is discretized temporally as
\begin{equation}
    \mathbf{f}\_{\Sigma}^{n+1} =  \sigma (\kappa_{\Sigma} \mathbf{n}_\Sigma)^n \delta_\Sigma  + \eta_\Sigma \left[ \Delta_\Sigma (\mathbf{u}) \right] ^{n+1},
    \label{eq:euler_fsigma}
\end{equation}
where $n$ and $n+1$ represent consecutive time steps.
As shown in~\cite{tolle_saample:_2020}, the Laplace-Beltrami operator is not implemented fully implicitly in OpenFOAM, and is instead represented in a semi-implicit way in~\cite{tolle_saample:_2020} with
\begin{equation}
    \Delta_\Sigma \mathbf{u} = \Delta \mathbf{u} 
    - \nabla \cdot \left[ (\mathbf{n}_\Sigma \cdot \nabla \mathbf{u} ) \otimes \mathbf{n}_\Sigma \right]
    - \kappa \left[ \left( \nabla \mathbf{u}  - (\mathbf{n}_\Sigma \cdot \nabla \mathbf{u}) \otimes \mathbf{n}_\Sigma \right) \cdot \mathbf{n}_\Sigma \right],
    \label{eq:laplace-beltrami}
\end{equation}
where the first term on the right hand side is treated implicitly and the remaining ones are represented explicitly.
In the spatial discretization, Denner et al. \cite{denner_artificial_2017} propose a normalization of the artificial viscosity with the mesh spacing. 
In the present model, normalization is performed by specifying the value of $\eta\_{\Sigma , c}$ in each cell center as
\begin{equation} 
    \eta_{\Sigma , c} = \mu_{\Sigma,c}  \, |\nabla \alpha_c| \, V_c^{1/3}
\end{equation}
where $\mu_{\Sigma,c}$ is the base viscosity value and $V_c$ is the cell volume.
The factor $|\nabla \alpha_c| \, V_c^{1/3}$ can be interpreted as a normalized filter field that activates the viscosity in the cells in the vicinity of the interface.
The normalization choice leads to $\eta_{\Sigma,c} \leq \SI{0.5}{} \mu_{\Sigma,c}$ for $3D$ Cartesian meshes. 
For the base viscosity~$\mu_{\Sigma,c}$, both the Raessi model $\mu\_{\Sigma,c, \mathrm{Raessi}} = \sigma \Delta t \, V_c^{-1/3}$ (cf. \cref{eq:implicit_st}) and a specified value $\mu_{\Sigma,c} =\mu_{\Sigma} = const $ are implemented.
%
The addition of~\cref{eq:euler_fsigma} to the momentum equation, \cref{eq:momentum}, is implemented within an OpenFOAM fvOption.
This allows for the selection of the method and its parameters at run time.
\notes{
\begin{itemize}
    \item mention that capillary time step is not totally mitigated because it is not implemented fully implicitly?
    \item mention whether interface is reconstructed before / after
    \item check that in my imlpementation the Raessi approach is correctly noramlized!
\end{itemize}
}

\section{Validation}
\label{sec:validation}
\subsection{Hydrodynamic benchmarks}
\label{subsec:validation-hydrodynamic}
For validation of the artificial viscosity model, two hydrodynamic benchmark cases are considered, specifically a 2D capillary wave and a 3D translating droplet.
\subsubsection{2D capillary wave}
\label{subsubsec:validation-hydrodynamic-wave}
%
\begin{table}
    \footnotesize
    \caption{Fluid parameters and surface tension coefficients. Left: Fluid parameters for all fluids considered in the validation cases. Right: Surface tension coefficients. Surface tension coefficient of Tween-80\texttrademark - air was measured in-house.\notes{todo - check}}
    \label{tab:05_fluid_params}
    \begin{minipage}{.5\linewidth}
      \centering
        \begin{tabular}{l c c}
            \toprule[1.5pt]
             & $\rho [\SI{}{kg/m^3}] $ & $\mu [\SI{}{m\pascal \second}]$ \\
            \midrule[0.5pt]
            Fluid \cite{denner_artificial_2017} & $1$ & $\SI{1.6394}{}$ \\
            Air \cite{VDI2010}& $1.19$ & $\SI{0.0182}{}$  \\             
            Water \cite{Lemmon2022, VDI2010}& $998.2$ &$\SI{0.9982}{}$  \\
            Oil \cite{Novec2022}& $1614.0$ & $\SI{1.2428}{}$ \\
            Tween$^{\tiny{\text{\textregistered}}}$-80~solution \cite{szymczyk_aggregation_2016}& $998.2$ &$\SI{0.9982}{}$  \\
            \bottomrule[1.5pt]
        \end{tabular}
    \end{minipage}%
    \begin{minipage}{.5\linewidth}
      \centering
        \begin{tabular}{l c c}
            \toprule[1.5pt]
             & $\sigma [\SI{}{mN/m}] $ \\
            \midrule[0.5pt]
            Fluid - fluid \cite{denner_artificial_2017} &   $\SI{250}{} \pi^{-3}$ \\
            Water - air \cite{IAPWS2014} & $\SI{72.74}{}$ \\
            Oil - water \cite{Brosseau2014}& $\SI{49.50}{}$\\
            Tween$^{\tiny{\text{\textregistered}}}$-80~solution - air & $\SI{35.00}{}$ \\
            & \\
            \bottomrule[1.5pt]
        \end{tabular}
    \end{minipage} 
\end{table}
\begin{figure}[!t]
    \includegraphics[width=0.49\textwidth]{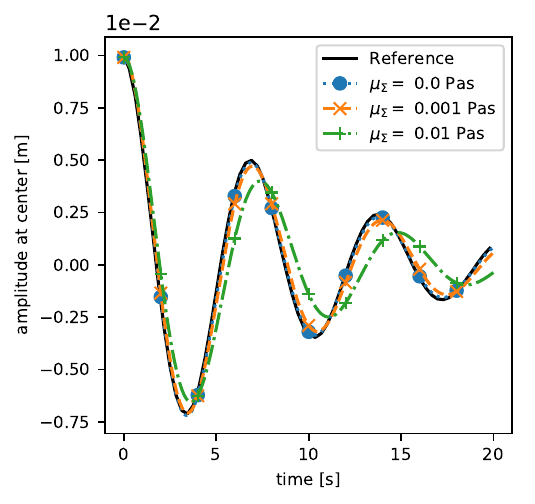}
    \hfill
    \includegraphics[width=0.49\textwidth]{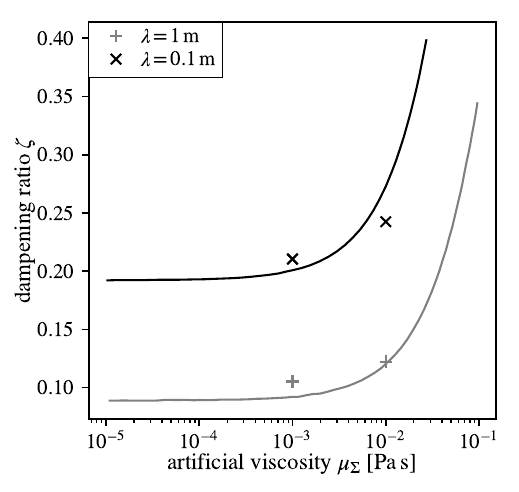}
     \caption{Results for the 2D capillary wave with interface viscosity. Left: Temporal evolution of the oscillation amplitude at the center for $\lambda = \SI{1}{m}$ with different values of interface viscosity. 
     Right: Dampening ratio for different wavelengths and viscosity values (symbols), compared to the solution from~\cite{denner_artificial_2017} (solid lines).}
     \label{fig:05_denner_avistudy}   
\end{figure}
\paragraph{Setup}
The decay of a capillary wave in a 2D domain is considered as given by Popinet~\cite{popinet_accurate_2009} and Denner et al. \cite{denner_artificial_2017}.
Two identical fluids are separated by an interface initialized with a sinusoidal wave shape with wavelength~$\lambda$ and initial amplitude~$0.01\lambda$.
Under the assumption that  gravitational forces can be neglected, the interface movement is solely induced by surface tension.
The fluid properties and surface tension coefficient taken from~\cite{denner_artificial_2017} are given in~\cref{tab:05_fluid_params}.
The domain has the size $\lambda \times 3\lambda$ and is discretized with $\Delta x = \lambda/32$ in an equidistant mesh.
All boundaries are treated as free-slip walls.
Two different wave lengths are considered, $\lambda =  \SI{1}{m}$ and  $\lambda =  \SI{0.1}{m}$, with respective temporal discretizations of $\Delta t = \SI{1e-3}{s}$ and $\Delta t = \SI{3.162e-5}{s}$ as given in~\cite{denner_artificial_2017}.
The artificial viscosity value is chosen as $\mu_\Sigma = 0$, $\mu_\Sigma = \SI{1e-3}{\pascal \second}$, and $\mu_\Sigma = \SI{1e-2}{\pascal \second}$.
%
\paragraph{Results}
The temporal evolution of the oscillation amplitude at the center of the interface is displayed on the left in~\cref{fig:05_denner_avistudy} for the capillary wave with wavelength $\lambda = \SI{1}{m}$.
The solution given by Prosperetti~\cite{Prosperetti1981} is given as a reference.
Only results of the solver interFlow-fPa are shown here. 
The same simulations with interIsoFoam yield very similar results which can be found in~\cref{app:denner_interIso}.
The influence of the artificial viscosity on the capillary wave oscillation matches the results shown by Denner et al.~\cite{denner_artificial_2017}.
Without the artificial viscosity model, the solution is in very good agreement with the analytic solution from~\cite{Prosperetti1981}.
Due to the additional dissipation introduced in the vicinity of the interface, the oscillation amplitude is reduced compared to the analytic solution.
Increasing the artificial viscosity results in stronger dampening. 
Furthermore, an increasing phase shift in the oscillation with higher artificial viscosity values is induced by the artificial dissipation.
%
The right subfigure in~\cref{fig:05_denner_avistudy} displays the dampening ratio of the capillary waves 
\begin{equation}
\zeta = \dfrac{ln(|A\_0|/|A\_1|)}{(\omega\_\sigma (t\_1 - t\_0))}    
\label{eq:05_dampening_ratio}
\end{equation}
for the wavelengths  $\lambda = \SI{1}{m}$ and $\lambda = \SI{0.1}{m}$ and artificial viscosity values $\mu_\Sigma = \SI{1e-3}{\pascal \second}$ and $\mu_\Sigma = \SI{1e-2}{\pascal \second}$.
In~\cref{eq:05_dampening_ratio}
\begin{equation}
    \omega\_\sigma = \sqrt{\dfrac{(\sigma  ({2 \pi }/{\lambda})^3)}{(\rho^- + \rho^+)} }
\end{equation}
is the capillary wave frequency, and $A\_0$, $A\_1$, and $t\_0$, $t\_1$ are two subsequent amplitudes and the corresponding points in time, respectively.
The results are compared to the curve proposed in~\cite{denner_artificial_2017} and show good agreement.
The dampening ratio for low artificial viscosity values is larger compared to \cite{denner_artificial_2017}, and the increase in dampening ratio with increasing artificial viscosity is less pronounced than seen in the reference. \notes{why? does this need more explanation?}
%
\begin{figure}[!t]
    \includegraphics[width=0.49\textwidth]{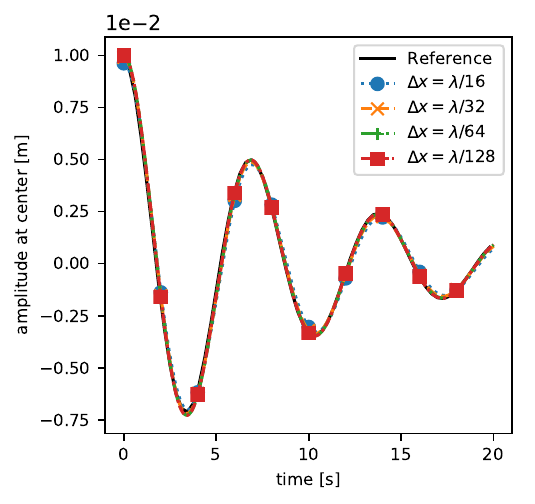}
    \hfill
    \includegraphics[width=0.49\textwidth]{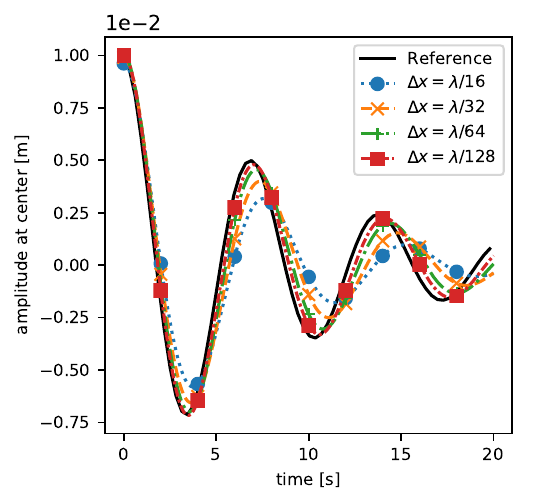}
     \caption{Results for mesh study of 2D capillary wave oscillation amplitude without interface viscosity and with interface viscosity. Left: results without interface viscosity for different mesh resolutions, denoted by the number of cells per wavelength. Right: mesh study with interface viscosity $\mu_\Sigma = \SI{0.01}{Pa s}$. interFlow-fPa is used.}
     \label{fig:05_denner_meshStudy}   
\end{figure}
%
\par The mesh dependence of the influence of $\mu_\Sigma$ is shown in~\cref{fig:05_denner_meshStudy}.
The temporal evolution of the oscillation amplitude for different mesh resolutions is presented.
As in~\cite{denner_artificial_2017}, four resolutions are chosen between $16$ and $128$ cells per wavelength $\lambda$.
The left subfigure displays the results for $\mu_\Sigma=0$ which show little mesh dependence.
In contrast, for an artificial viscosity value of $\mu_\Sigma=\SI{0.01}{\pascal \second}$, the mesh dependence is more pronounced, as shown in the right subfigure of~\cref{fig:05_denner_meshStudy}.
The dampening effect of the artificial viscosity increases for lower resolutions, matching the findings of~\cite{denner_artificial_2017}.
This can be attributed to the artificial viscous term being active exclusively in the cells surrounding the interface.
Therefore, the volumetric region of influence of the artificial viscosity increases with increasing cell size.
\notes{
\begin{itemize}
    \item mention this? Compared to Denner et al., a larger impact of $\mu_\Sigma$ on the amplitude is seen in the present study. We cannot confirm a stronger dampening effect of the artificial viscosity model on the shorter wave ($\lambda = \SI{0.1}{m}$). \notes{why?}
    \item Compared to the physical viscosity in the domain, this amounts to ??
\end{itemize}
}
\subsubsection{3D translating droplet}
\label{subsubsec:validation-hydrodynamic-translatingDroplet3D}
\begin{figure}[!t]
    \includegraphics[width=0.49\textwidth]{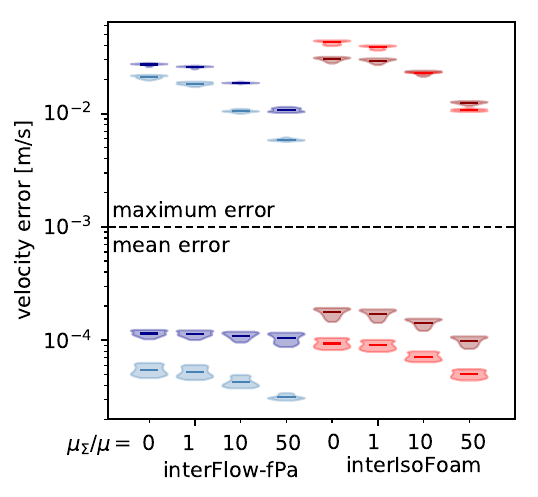}
    \hfill
    \includegraphics[width=0.49\textwidth]{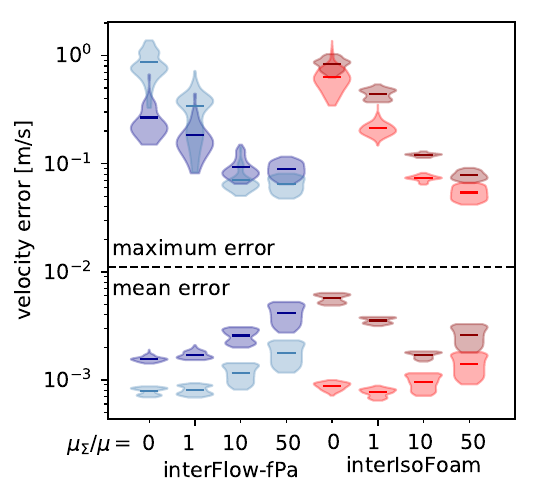}
    \caption{Results for 3D translating droplet. Maximum and mean velocity errors for resolution 32 (dark colors) and resolution 64 (light colors) with different values for $\mu_\Sigma$. Left: fluid pairing oil-water, right: fluid pairing water-air.}
     \label{fig:05_transDroplet3D}   
\end{figure}
\paragraph{Setup}
As a 3D hydrodynamic benchmark case, a translating droplet transported through the domain by a flow velocity $\mathbf{U}=(1,0,0) \, \SI{}{m/s}$ is considered.
The case is described in full detail by Lippert et al.~\cite{lippert_benchmark_2022}.
Here, the same setup as in~\cite{lippert_benchmark_2022} is used, and a range of artificial viscosities is applied.
Two fluid pairings are considered, namely a water droplet in air and an oil droplet in water. 
The corresponding density, viscosity, and surface tension values are given in~\cref{tab:05_fluid_params}.
The values of the artificial viscosity are chosen based on the physical viscosities in the domain.
The larger dynamic viscosity in the domain, called $\mu$ in the following for simplicity, is chosen as the base value. The artificial viscosity is chosen as $\mu_\Sigma = 0$,  $\mu_\Sigma = 1 \mu$,  $\mu_\Sigma = 10 \mu$ and $\mu_\Sigma = 50 \mu$.
Two spatial resolutions of the domain are considered, namely $32$ and $64$ cells in streamwise direction.
\paragraph{Results}
As described in~\cite{lippert_benchmark_2022}, two error metrics are considered for evaluation, namely the maximum velocity error
\begin{equation}
    L_\infty(\mathbf{u}) = \mathrm{max} \left( | \mathbf{u}_i - \mathbf{U} | \right)
\end{equation}
and the mean velocity error
\begin{equation}
    L_1(\mathbf{u}) = \dfrac{1}{N\_{cells}} \sum^{N\_{cells}}_i | \mathbf{u}_i - \mathbf{U} | .
\end{equation}
The resulting maximum and mean velocity errors as a function of the artificial viscosity are given as violin plots in~\cref{fig:05_transDroplet3D}.
The results for the fluid pairing oil-water and water-air are displayed in the left and right subfigure, respectively. 
Without artificial viscosity, the results match the data given by Lippert et al.~\cite{lippert_benchmark_2022}.\notes{double check!}
In all cases, the magnitude of the maximum velocity error decreases with increasing artificial viscosity.
This effect appears in the solution with the coarse mesh, displayed with dark colors, as well as with the fine mesh, displayed in light colors.
The mean velocity errors are matching the behavior of the maximum errors for the oil-water fluid pairing.
In the water-air case, however, the mean error rises with increasing artificial viscosity.
This can be attributed to an overpronounced dampening effect which reduces the velocity in the domain surrounding the interface.
The results indicate that spurious currents can be effectively dampened with the artificial viscosity model for both considered solvers.
However, the effect of the artificial viscosity is sensitive to the physical properties of the considered fluids.
Excessive artificial viscosities can, although reducing the maximum velocity error in the domain, increase the mean error by overdampening.
\notes{
\begin{itemize}
    \item double check color coding
    \item mention that it is plotted for $t>0.5 tmax$
    \item low viscosities, highest density ratio and largest surface tension coefficient
\end{itemize}
}
%
\subsection{Wetting benchmarks}
\label{subsec:validation-wetting}
\begin{figure}[!t]
    \centering
    \includegraphics[width=0.99\textwidth]{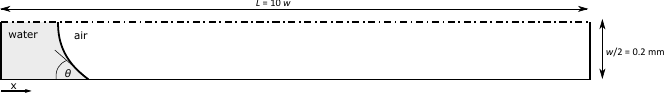}
    \caption{Geometry sketch for wetting benchmark cases.}
    \label{fig:05_forcedChannelWetting_setup}
\end{figure}
\paragraph{General setup information}
To validate the artificial interface viscosity for wetting in regimes relevant for LoC applications, two 2D capillary wetting cases are considered, only differing in their inlet and outlet boundary conditions.
The domain is depicted in~\cref{fig:05_forcedChannelWetting_setup}.
A symmetry plane is introduced at the top boundary by employing a free-slip velocity condition, a zero-gradient pressure condition and a constant contact angle of $\theta=\SI{90}{\degree}$.
At the bottom wall, the contact angle is chosen as $\theta=\SI{45}{\degree}$, and a partial slip condition is applied for the velocity with a slip length arbitrarily chosen as one cell height. \notes{todo - explain, use partial slip.} 
Only the fluid pairing water-air is considered for simplicity.
A uniform mesh is created with blockMesh with discretization length $\Delta x = w/32$, where $w$ is the width of the capillary.
As time step, $\Delta t = \SI{1e-6}{s}$ is chosen, which satisfies both the capillary time step constraint and the CFL condition.\notes{todo explain more for all cases? write down all delta ts? in table?}
Three values of the artificial viscosity are considered, namely $\mu_\Sigma = 0$,  $\mu_\Sigma = 10 \mu$ and $\mu_\Sigma = 100 \mu$, where $\mu$ is the dynamic viscosity of water.
\notes{
\begin{itemize}
    \item initialization of the interface with cylinder
\end{itemize}
}
\subsubsection{Capillary rise}
\label{subsubsec:validation-wetting-capillaryRise}
\begin{figure}[!tb]
    \includegraphics[width=0.445\textwidth]{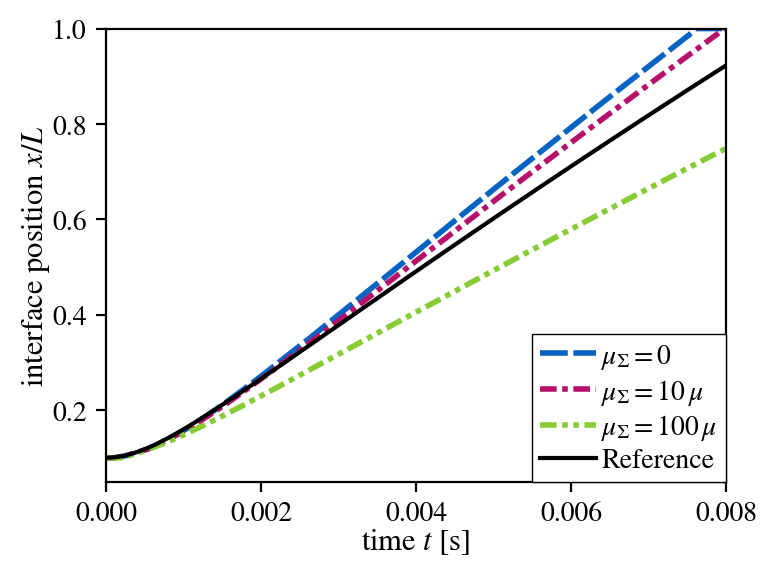}
    \hfill
    \includegraphics[width=0.545\textwidth]{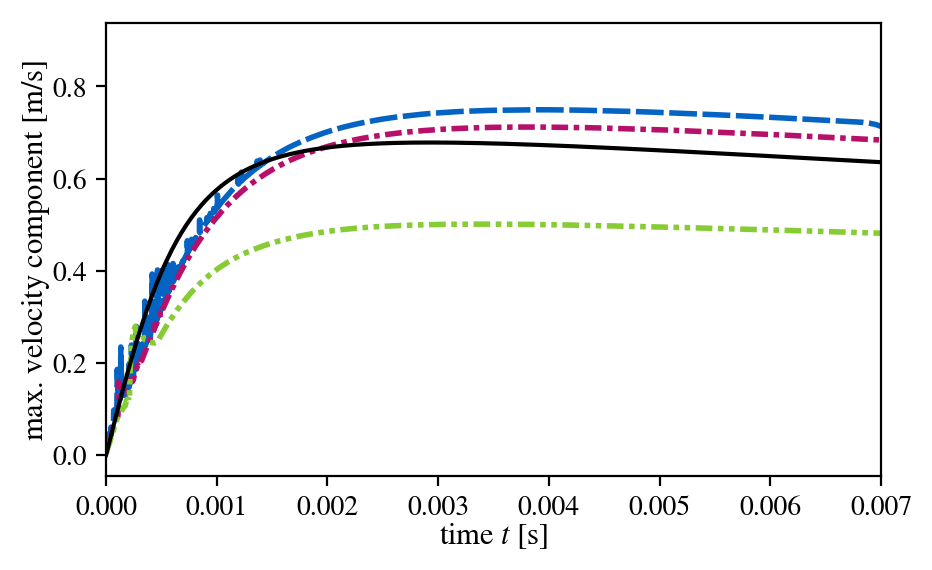}
     \caption{Results for the 2D horizontal capillary rise with interFlow-fPa. Left: temporal progression of the meniscus position compared to the analytical solution~\cref{eq:05_bosanquet_solution}. Right: Temporal evolution of the maximum velocity component in the domain compared to the temporal derivative of~\cref{eq:05_bosanquet_solution}.}
     \label{fig:05_bosanquet_interFlo}   
\end{figure}

\begin{figure}[!tb]
    \includegraphics[width=0.445\textwidth]{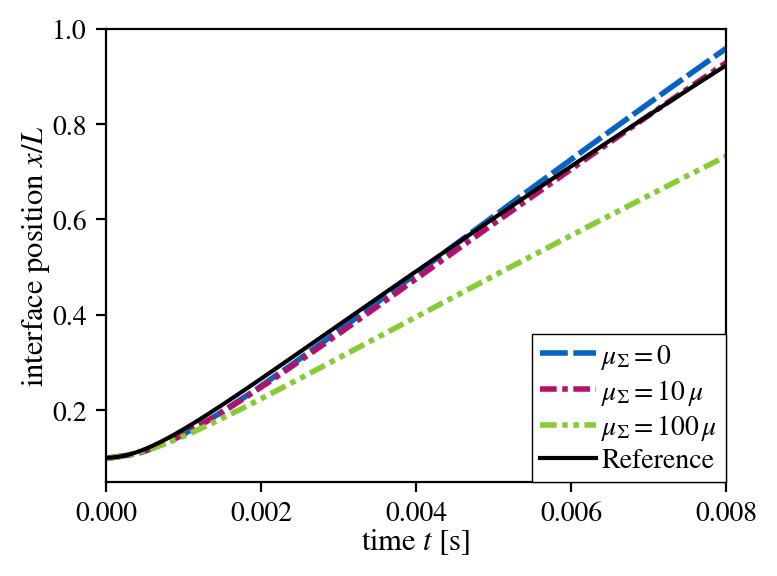}
    \hfill
    \includegraphics[width=0.545\textwidth]{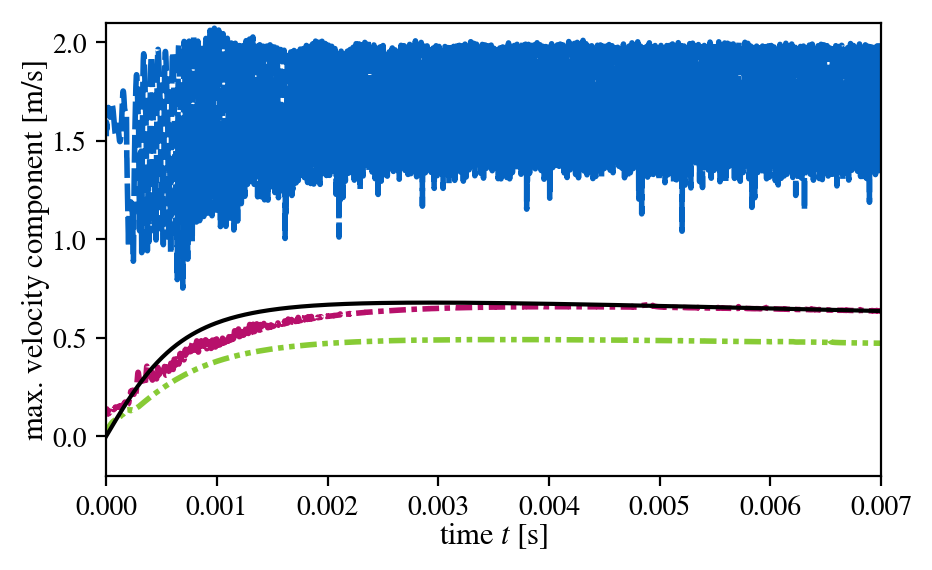}
     \caption{Results for the 2D horizontal capillary rise with interIsoFoam. Left: temporal progression of the meniscus position compared to the analytical solution~\cref{eq:05_bosanquet_solution}. Right: Temporal evolution of the maximum velocity component in the domain compared to the temporal derivative of~\cref{eq:05_bosanquet_solution}.}
     \label{fig:05_bosanquet_interIso}   
\end{figure}

\paragraph{Setup}
A 2D horizontal capillary rise is considered as a standard wetting benchmark.
As analytical solution, Bosanquet~\cite{bosanquet_lv._1923} proposes a model for the meniscus center progression $x(t)$ considering viscous, capillary and inertial forces. 
The corresponding differential equation reads
%
\begin{equation}
\label{eq:05_bosanquet_eq}
\frac{d}{dt} (x \dot{x })  + a x \dot{x} = b. 
\end{equation}
In the case of a capillary rise between two parallel plates, the parameters $a$ and $b$ are
\begin{align}
\label{eq:05_bosanquet_params}
	a & = \dfrac{12 \mu}{\rho w^2} \\
	b & = \frac{1}{\rho} \left( \dfrac{2 \sigma \cos(\theta)}{w} \right).
\end{align}
Here, $w$ is the width of the capillary, and the density and viscosity are those of the liquid, while the gas phase is neglected.
A parabolic flow profile is assumed throughout the domain.
The solution to~\cref{eq:05_bosanquet_eq} is
\begin{equation}
\label{eq:05_bosanquet_solution}
    x(t) = +\sqrt{\left( x(t=0)\right) ^2 + \dfrac{2b}{a} t - \dfrac{2b}{a^2}  (1 - e^{-at})} \ 
\end{equation}
which is used as a reference for the simulations.\notes{references!!!!}
%
Zero total pressure boundary conditions are applied at the inlet and outlet.
\notes{
\begin{itemize}
    \item goal: validate AVI for standard wetting benchmark within regime relevant for LoC
    \item (what I will not consider: slip length study ( maybe in appendix / additional data repo), contact angle study
\end{itemize}
}
%
\paragraph{Results}
On the left side of~\cref{fig:05_bosanquet_interFlo}, the progression of the meniscus position $x(t)$ along the symmetry plane is displayed for the solver interFlow-fPa.
The analytical solution given in~\cref{eq:05_bosanquet_solution} serves as a reference.
The maximum velocity component in the domain is given on the right side of~\cref{fig:05_bosanquet_interFlo}.
As reference, the expected maximum velocity component in the domain is approximated with $1.5\, dx/dt$ and also given in the figure.
With $\mu_\Sigma = 0$ and  $\mu_\Sigma = 10 \mu$, the analytical capillary rise velocity is exceeded shortly after initialization.
With $\mu_\Sigma = 100 \mu$, the increased dampening in the vicinity of the interface reduces the velocity of the meniscus significantly.
Except for small oscillations in the first time steps, the velocity is not visibly obscured by spurious currents in either of the cases.
This contrasts the results obtained with interIsoFoam shown in ~\cref{fig:05_bosanquet_interIso}.
While the meniscus progression is predicted close to the analytical solution for $\mu_\Sigma = 0$, the maximum velocity component differs significantly from the analytical curve, oscillating strongly and reaching values of more than double the interface speed.
The high velocity is  attributed to spurious currents appearing in the vicinity of the interface.
With increased artificial viscosity, the maximum velocity component is reduced.
Parallel to the interFlow-fPa case, the high viscosity value of $\mu_\Sigma = 100 \mu$ results in overdampening such that the interface velocity is underpredicted.
\par The results show that the artificial viscosity model is able to dampen spurious currents in the vicinity of the interface in the capillary rise case, especially for interIsoFoam where spurious currents are more prevalent.
%
\subsubsection{Forced channel wetting}
\label{subsubsec:validation-wetting-forced}
\paragraph{Setup}
A 2D horizontal capillary with a given block velocity $U=\SI{0.01}{m/s}$ at the inlet is considered.
At the outlet, the pressure is set to zero and a zero-gradient condition is chosen for the velocity.
The remaining setup details are given in the first paragraph of~\cref{subsec:validation-wetting} and in the sketch in~\cref{fig:05_forcedChannelWetting_setup}.
\begin{figure}[!t]
    \centering
    \includegraphics[width=0.95\textwidth]{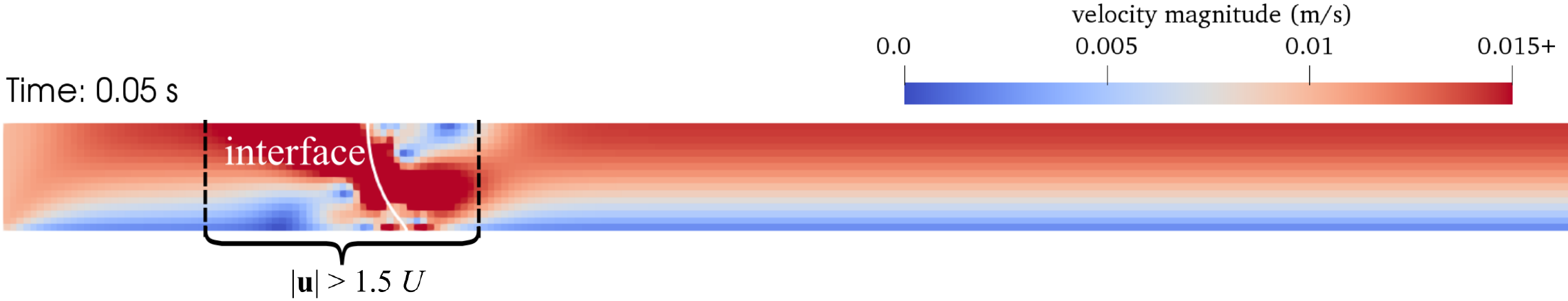}
    \caption{Schematic view of the region of influence (ROI) of the spurious currents in the forced wetting case with interIsoFoam and $\mu_\Sigma = 10 \mu$.}
    \label{fig:05_forcedChannelWetting_roi}
\end{figure}
\begin{figure}[!t]
    \centering
    \includegraphics[width=0.49\textwidth]{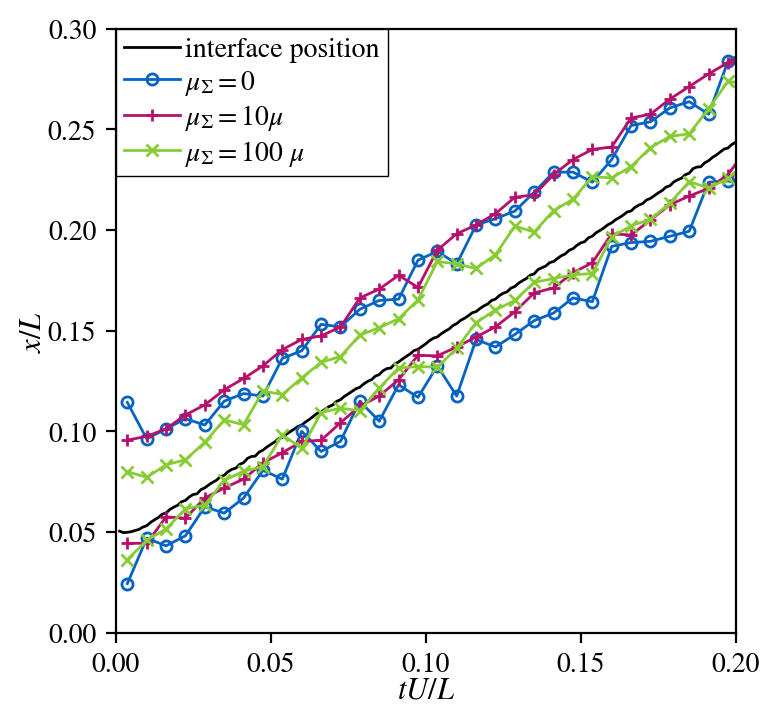}
    \hfill
    \includegraphics[width=0.49\textwidth]{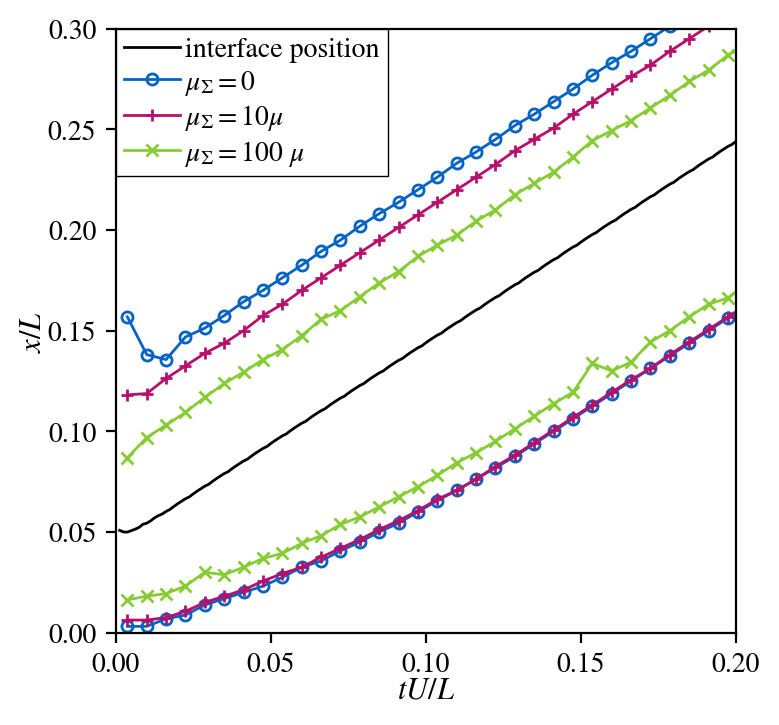}
    \caption{Temporal progression of the meniscus and the minimum and maximum position of the ROI of spurious currents for the 2D forced channel wetting case. Left: interFlow-fPa, right: interIsoFoam.}
    \label{fig:05_forcedChannelWetting_results}
\end{figure}
\paragraph{Results}
In the evaluation of the simulations, the magnitude of the maximum velocity is not considered.
Instead, the aim is to estimate the region of influence of the spurious currents occurring in the vicinity of the interface.
The region of influence (ROI) of the spurious currents is defined as the region where $|\mathbf{u}|/U \geq \SI{1.5}{} $, as schematically shown in~\cref{fig:05_forcedChannelWetting_roi}.
In ~\cref{fig:05_forcedChannelWetting_results}, the temporal evolution of the ROI is displayed for both solvers.
Along with the maximum and minimum position of the ROI, the progression of the meniscus along the symmetry plane $x(t)$ is shown.
The normalized time $tU/L$ and normalized position $x(t)/L$ are displayed, where $L$ is the channel length.
The left subfigure shows the results obtained with interFlow-fPa.
In contrast to the capillary rise case presented in~\cref{subsubsec:validation-wetting-capillaryRise}, the meniscus progression is forced by the given inflow velocity, therefore not being visibly influenced by the artificial viscosity.
The ROI spans approximately five percent of the channel length without artificial viscosity applied.
In the region downstream of the meniscus, the ROI is larger than the ROI upstream of the meniscus.
This can be attributed to the fluid pairing.
The air phase is located downstream of the meniscus, and the small dynamic viscosity compared to water is assumed to be the cause for the comparably larger influence of spurious currents.\todo{explain more}
The application of the artificial viscosity model with $\mu_\Sigma = 10 \mu$ does not significantly improve the results.
However, with the artificial viscosity increasing to $\mu_\Sigma = 100 \mu$ , the ROI is reduced.
On the air side (downstream of the meniscus), the ROI is approximately halved, while on the water side (upstream of the meniscus), the ROI almost disappears completely.
A similar trend is visible in the results obtained with interIsoFoam, displayed in the right subfigure of~\cref{fig:05_forcedChannelWetting_results}.
The application of the artificial viscosity reduces the ROI on both sides of the meniscus.
However, with interIsoFoam, the magnitude of the ROI is larger than with interFlow-fPa, specifically on the water side.
In both cases, the artificial viscosity has a stronger effect on the ROI on the air side upstream of the meniscus, which can again be attributed to the fluid properties.
\par As a result of the validation with hydrodynamic and wetting cases, the solver interFlow-fPa is chosen for the following simulations due to its overall better performance.
\notes{
\begin{itemize}
    \item roi depends on inflow velocity, does AVI influence too?
    \item (go backwards from results I want to show in experimental validation section! if everything can be explained there, I don't need it here.) 
\end{itemize}
}

\section{Simulation and experimental investigation of interface traversing through microcavities}
\label{sec:results}
\begin{figure}[!tb]
    \centering
    \raisebox{-0.5\height}{\includegraphics[width=0.49\textwidth]{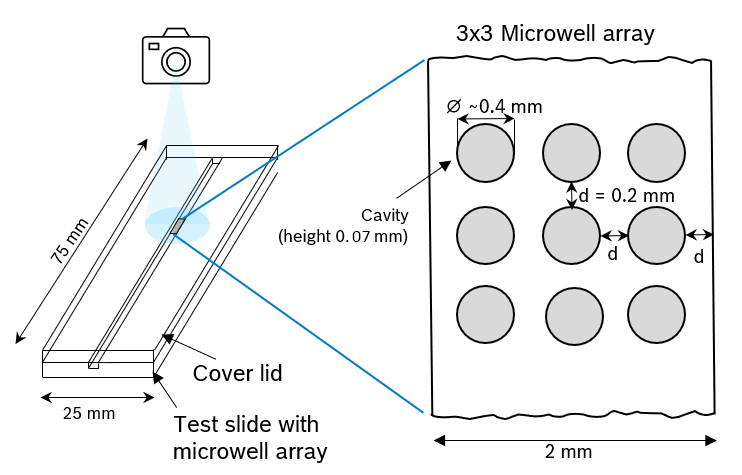}}
    \hfill
    \raisebox{-0.5\height}{\includegraphics[width=0.49\textwidth]{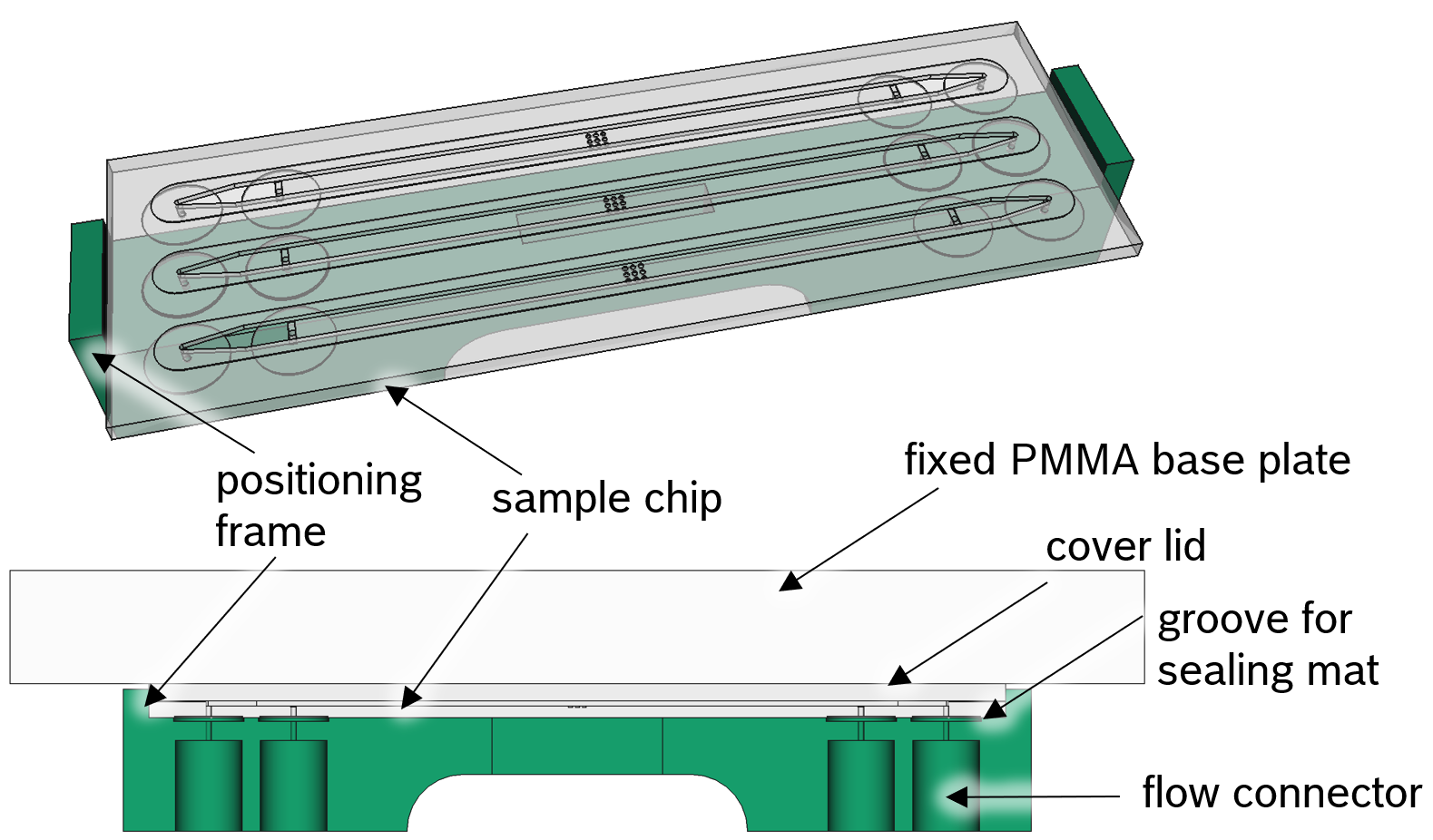}}
    \caption{Experimental setup. Left: Scheme and dimensions of the sample chip and microcavity array. Right: Micromilled chip and positioning frame (top), cut view of the assembled setup (bottom).
    \notes{to do - photo of full setup? arrows for flow inlet and outlet. rubber mats and tube connections. impove scheme.}}
    \label{fig:06_experimental_setup}
\end{figure}
\subsection{Setup}
\label{subsec:results-expValidationSetup}
\paragraph{Experimental setup}
An interfacial flow through a channel with microcavities is evaluated optically.
As test samples, injection-molded Polymethyl-methacrylate (PMMA) microscopy slides (Microfluidic ChipShop) are processed by micromilling to achieve the chip geometry shown in ~\cref{fig:06_experimental_setup} (right). 
Each chip contains three identical open-channel structures with a 3x3 microcavity setup in the channel center. 
All relevant measurements are given in~\cref{fig:06_experimental_setup}.
The channels are closed with a cover lid.
The assembly consisting of chip, cover lid and metal positioning frame is clamped below a thick PMMA base plane.
Fastening is achieved solely by clamping and without glue usage 
This allows for simple disassembly, cleaning, and reusage of the chips for reproducability tests, and for postprocessing investigations of the material after the test was performed.
Furthermore, the clamping ensures the absence of glue contamination at the plate interfaces.
For flow control, a syringe pump\notes{(syringe pump model?)} is employed.
The chip inlet and outlet are located at the channel ends.
Via a downwards tube connection through the metal holding frame, liquid is supplied at inlet and released at the outlet.
Rubber sealings at the liquid connectors provide leakage-free liquid transfer into the chip.
Due to the positioning of the cavity array in the channel center, influences of the inlet and outlet on the cross-cavity flow are negligible for low mass flow rates. \notes{(to do - proof? flow development length is not really useful for two-phase flows?)}
As test liquid, a $\SI{0.1}{\percent}$ Tween$^{\tiny{\text{\textregistered}}}$-80~aquaeous solution is used (Tween$^{\tiny{\text{\textregistered}}}$-80, Croda International plc). 
The fluid properties and surface tension coefficient with air are given in~\cref{tab:05_fluid_params}.
While the density and viscosity are very similar to water \cite{szymczyk_aggregation_2016}, the surfactant Tween$^{\tiny{\text{\textregistered}}}$-80~reduces the surface tension coefficient and the apparent contact angle on PMMA surfaces compared to pure water~\cite{rehman_surface_2017, szymczyk_effect_2018}.
It is used to improve the filling of the microcavities compared to pure water.
An RGB LED panel illuminates the chip using backlight technology (from below), and a high-speed camera\notes{camera model?} is employed to record the interface traversing through the microcavity array.
\notes{
	\begin{itemize}
		\item mention second connectors for pressure meaurements? or remove them from the sketches and not mention them?
        \item The experiments were repeated at least three times. 
        \item mention that the refractive index changes at the contact line
	\end{itemize}
}
\paragraph{Numerical setup}
In the simulation solely the cavity array is considered, inflow and outflow effects in the channel are neglected.
The geometry measurements are identical to the experimental geometry given in~\cref{fig:06_experimental_setup}.
The computational domain is halved by a symmetry plane using a free slip condition for the velocity, a zero-gradient condition for pressure and a static contact angle $\theta = \SI{90}{\degree}$.
At the inlet, a constant velocity block profile is given, and the outlet is defined as a zero pressure boundary.
All remaining domain boundaries are walls with a partial slip condition for the velocity, where the slip length is arbitrarily chosen as one cell height.
A constant static contact angle $\theta$ is defined at the walls which is defined in~\cref{subsec:results-contactAngle_velocitiy}.
The computational mesh is created with snappyHexMesh, based on a uniform block mesh with cell edge lengths  $\Delta x = \SI{2e-5}{m}$, resulting in a total number of $204000$ cells that are partitioned into four domains for parallel computing.
A constant time step of $\Delta t  = \SI{5e-6}{s}$ is used.\notes{todo -  get cfl and capillary dt for this.}
As solver, interFlow-fPa is employed due to its overall superior performance compared to interIsoFoam in the validation cases presented in~\cref{sec:validation}.
To study the influence of the artificial viscosity, the artificial viscosity is chosen as $\mu_\Sigma=\SI{0}{}$ and $\mu_\Sigma=\SI{10}{}\mu$. 
The wisp correction described in~\cref{subsec:wisp} is activated in selected cases with a wisp tolerance $\epsilon\_w=\SI{1e-5}{}$. 
\notes{
	\begin{itemize}
		\item mesh study results shortly / appendix?
        \item time step study shortly / appendix?
        \item explain how I determine whether cavities are filled?
        \item setup - CL initialization as cylinder with specified contact angle at side walls.
	\end{itemize}
}
\subsection{Contact angle calibration}
\label{subsec:results-contactAngle_velocitiy}
Prior to the study of the interface shape and movement in the cavity array, a constant contact angle $\theta$ must be determined as boundary condition for the simulations.
Therefore, the apparent contact angle in the channel is measured for the two different flow conditions.
This is achieved with image processing, using recordings of the contact line movement upstream the cavity array.
Five measurements of the contact angle at different points in time are averaged.
The measured contact angle is $\theta=\SI[separate-uncertainty = true]{105(2)}{\degree}$ for the case with $U=\SI{0.01}{m/s}$, and $\theta=\SI[separate-uncertainty = true]{90(2)}{\degree}$ for the case with $U=\SI{0.001}{m/s}$.
In the latter case, instead of the PMMA cover plate, a polycarbonate cover plate with increased roughness is used to improve the cavity filling process, and a contact angle of $\theta=\SI[separate-uncertainty = true]{120(5)}{\degree}$ is assumed at the upper channel boundary for the occuring velocity of $U=\SI{0.001}{m/s}$.
\notes{
    \begin{itemize}
		\item The roughness at the channel bottom is in order of magnitude of $\SI{5e-6}{m}$ due to the manufacturing.
        \item Increased roughness influences static contact angle and mostly dynamic contact angle
        \item show CA measurements - appendix?
	\end{itemize}
}
\subsection{Interface shape and movement comparison}	
\begin{figure}[!t]
    \centering
    \includegraphics[width=0.32\textwidth]{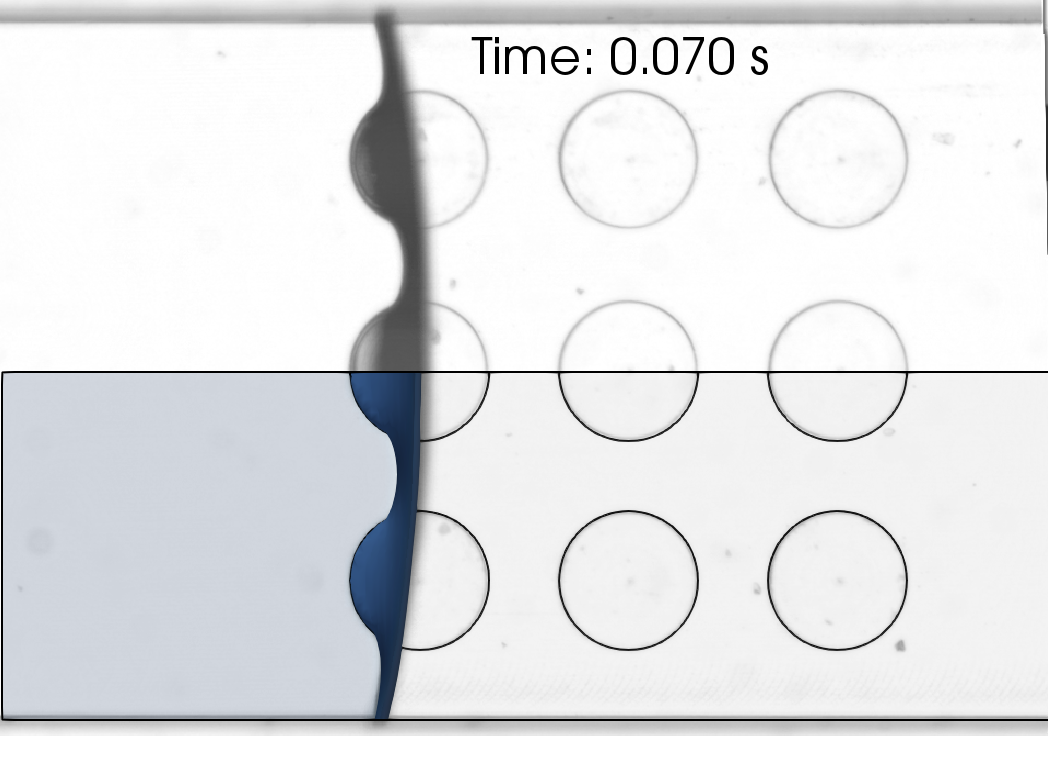}
    \hfill
    \includegraphics[width=0.32\textwidth]{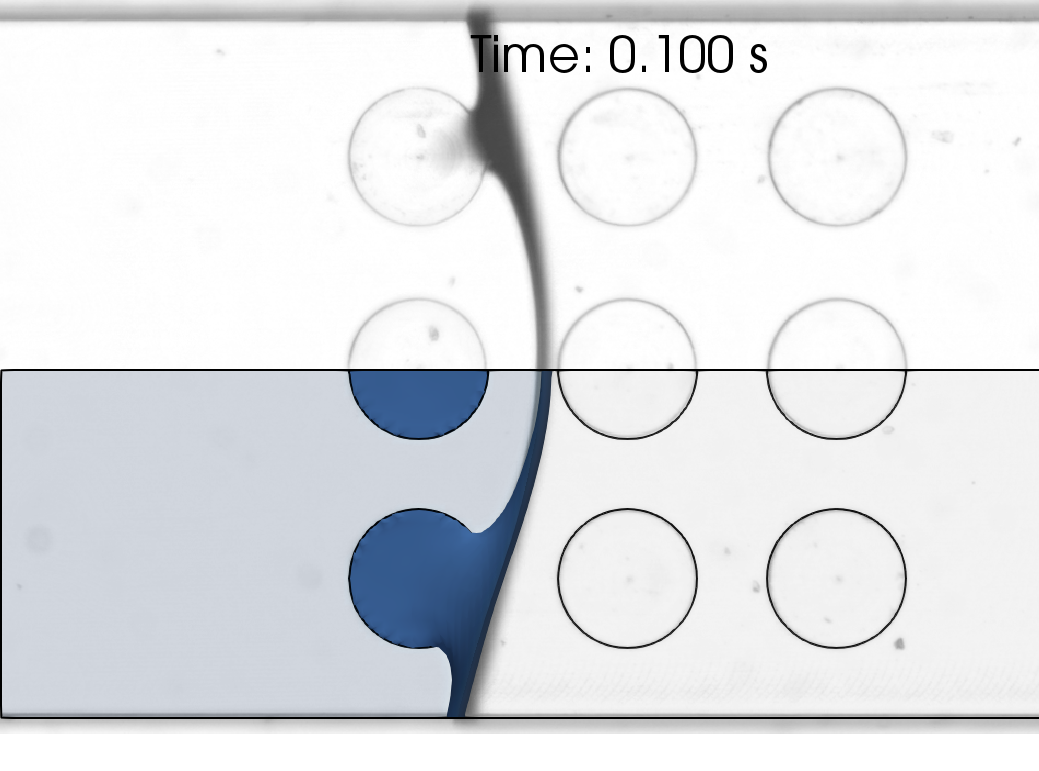}
    \hfill
    \includegraphics[width=0.32\textwidth]{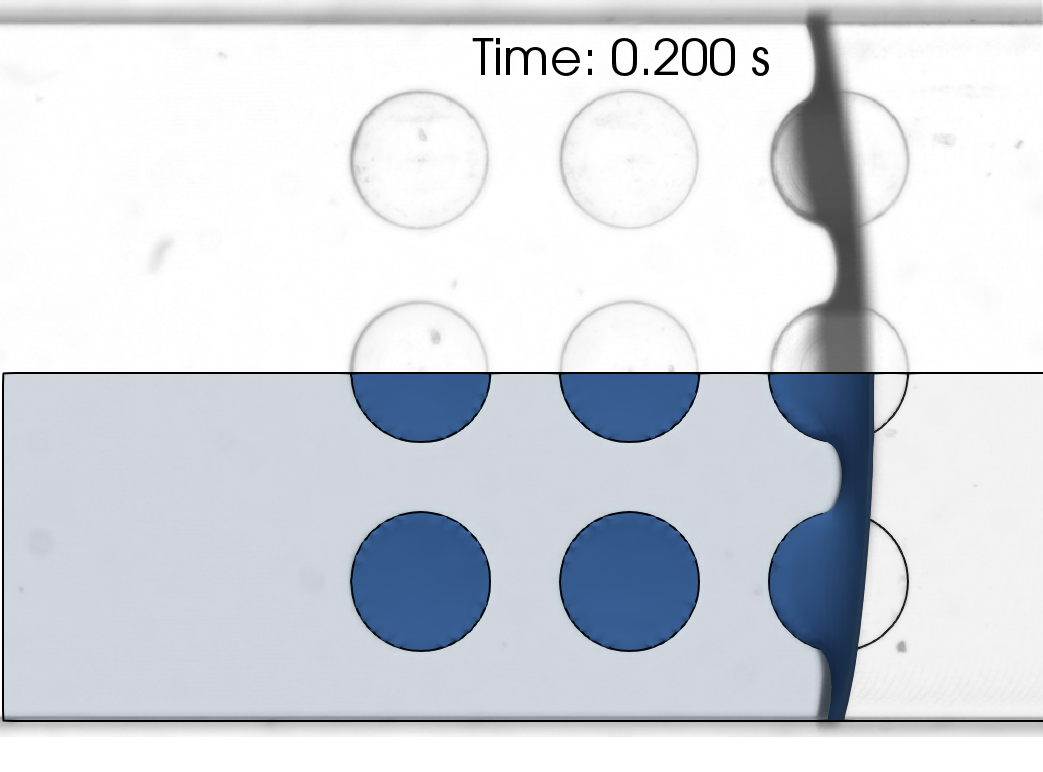}
    \includegraphics[width=\textwidth]{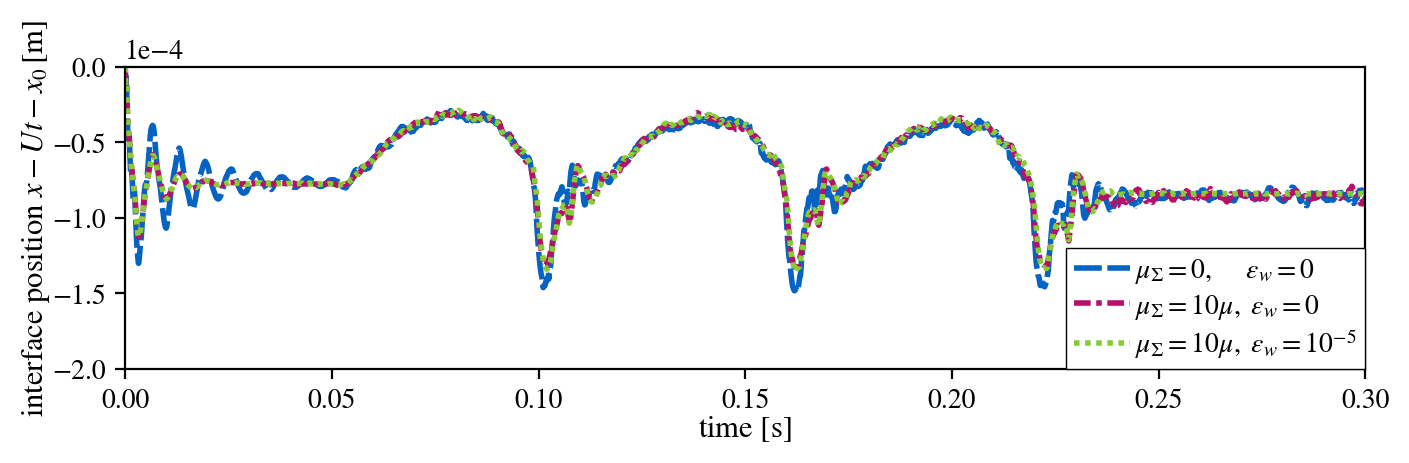}
    \includegraphics[width=\textwidth]{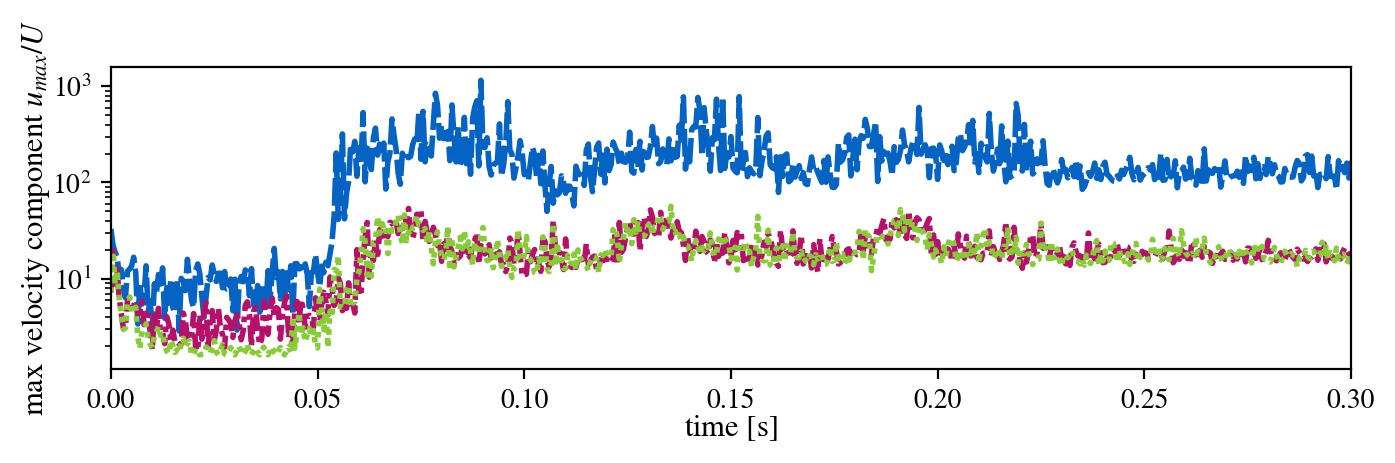}
    \caption{Results for 3x3 cavity flow with $U=\SI{0.01}{m/s}$ and contact angle $\theta=\SI{105}{\degree}$. Top: Visual comparison of experiment and simulation. Center: Interface progression for different simulation setups. Bottom: Maximum velocity component for different simulation setups. \notes{label images with simulation and experiment}}
    \label{fig:06_results_notfilling}
\end{figure}
\subsubsection{Traversing interface without cavity filling}
\label{subsubsec:06_notfilling}
For the first test case, the inlet velocity is chosen as $U=\SI{0.01}{m/s}$ and the corresponding contact angle at the wall is $\theta=\SI{105}{\degree}$.
The visual comparison of the experiment and the simulation with $\mu_\Sigma=0$ is given in the upper subfigure of~\cref{fig:06_results_notfilling} for three points in time.
In the experiments, the cavities are not filled, instead an air cushion rests inside the cavities.
The position of the interface in the third dimension, perpendicular to the camera view, cannot be determined directly in the experiments.
However, the visual assessment of the contact line movement shows that a pinning effect occurs at the cavity edge, which results in the contact line not moving into the cavity.
This is confirmed by the simulation.
Good agreement of the interface movement is achieved between the simulation and the experiment.
No difference is observed between the interface shape acquired with $\mu_\Sigma=0$ and $\mu_\Sigma=\SI{10}{}\mu$. 
\par The movement of the contact line in the simulation is further studied in the center subfigure in~\cref{fig:06_results_notfilling}.
The interface progression along a horizontal line at the side wall at half the channel height is shown.
To examine the interface progression $x(t)$ in detail, it is displayed in a moving reference frame as $x(t) - U t- x(t=0)$, where $U$ is the inflow velocity and $t$ is the current point in time.
After oscillations following the flow initialization, the contact line moves with constant velocity in the channel area upstream of the cavities.
Upon reaching a cavity row, the contact line is pinning at the cavity edges and is thus advancing faster at the channel wall due to the prescribed constant mass flow.
Once the interface surpasses a cavity row (e.g. at $t=\SI{0.1}{s}$), a jump in the meniscus progression at the wall occurs, followed by oscillations.
With the artificial viscosity model applied, the contact line movement does not change significantly, except for the oscillation amplitude being slightly reduced.
This indicates that the artificial viscosity value is small enough to maintain a suitable accuracy of the interface movement prediction.
Furthermore, the wisp removal algorithm with the chosen wisp tolerance $\epsilon\_w=\SI{1e-5}{}$ has no significant influence on the contact line position.
\notes{I am plotting $xm-(u_in*tm)-x0$ over $tm$}
\par The effect of the artificial viscosity model and wisp removal algorithm on spurious currents is displayed in the bottom subfigure of~\cref{fig:06_results_notfilling}.
The maximum velocity component in the domain, normalized with the inlet velocity, is depicted.
Strikingly, with no artificial viscosity applied, the maximum velocity exceeds the inlet velocity by up to two orders of magnitude before the contact line reaches the cavities.
Upon the contact line entering the vicinity of the cavity array, the maximum velocity rises, exceeding the inlet velocity by up to three orders of magnitude.
The high velocities can be attributed to spurious currents, which grow excessively large when the interface is located at a sharp cavity edge.\notes{more explanations here?}
With an applied artificial viscosity of $\mu_\Sigma=\SI{10}{}\mu$, the magnitude of the spurious currents is drastically reduced by over one order of magnitude.
With active wisp correction, a further reduction in the maximum velocity can be achieved in the first stage of the flow before the interface reaches the cavities.\notes{explain more?}
The mass loss due to the wisp removal remains below~$m < \SI{0.005}{} V \rho\_{air}$, with $V$ being the domain volume.
%
\begin{figure}[!t]
    \centering
    \includegraphics[width=0.32\textwidth]{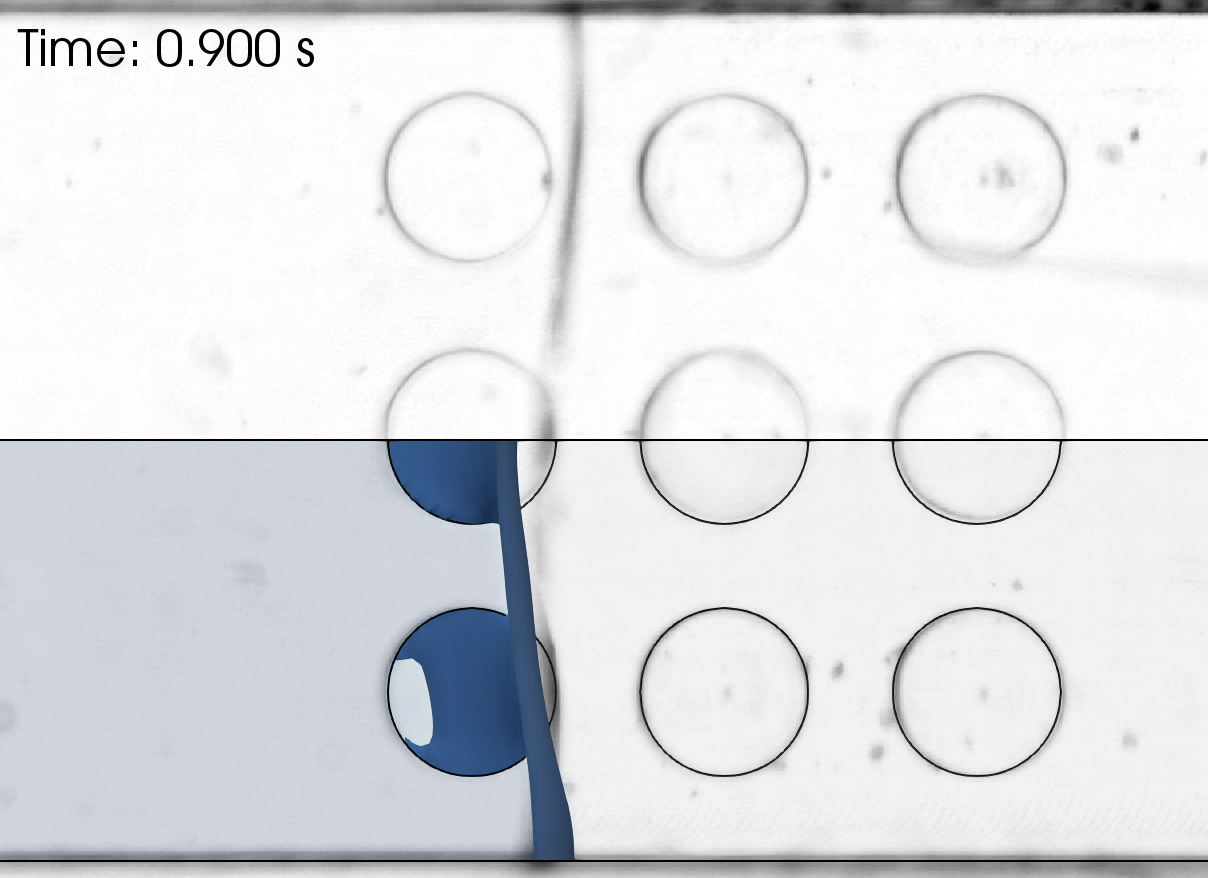}
    \hfill
    \includegraphics[width=0.32\textwidth]{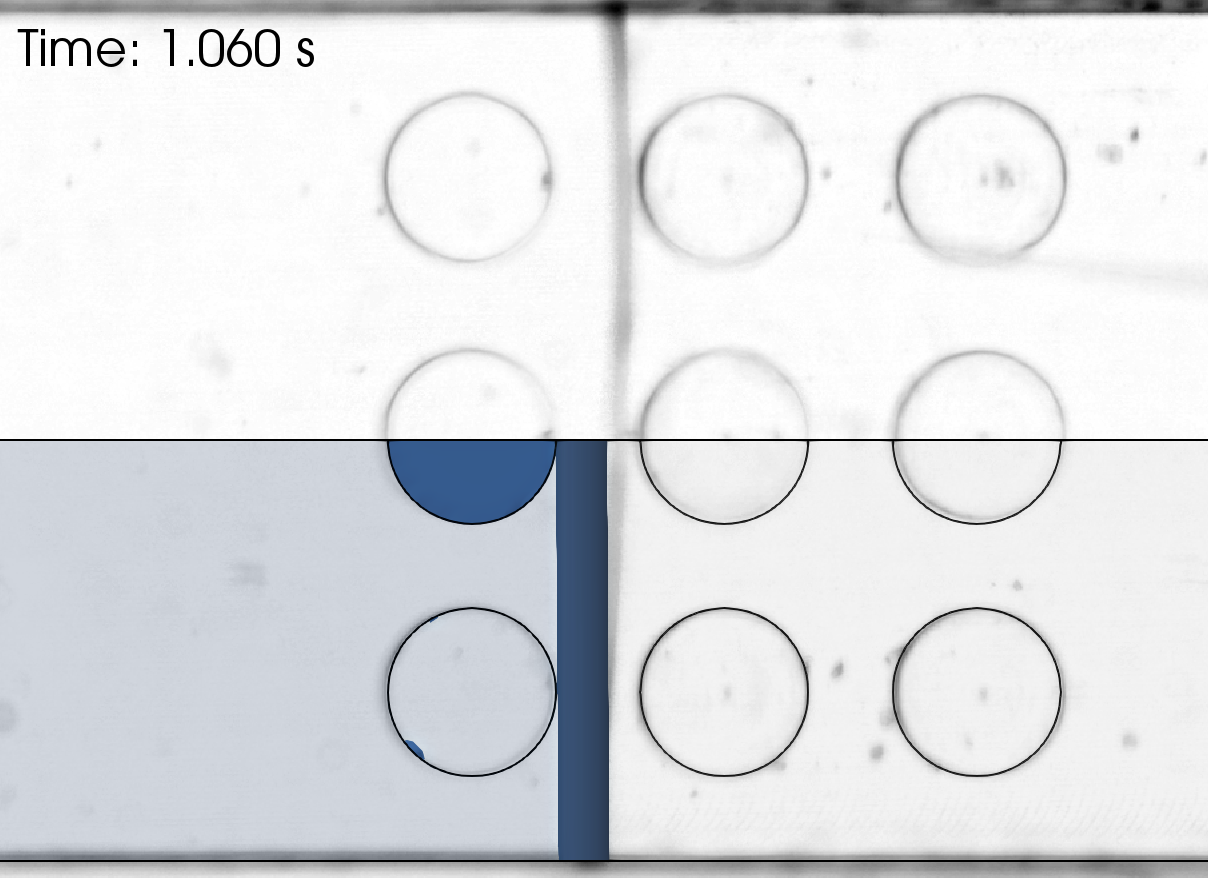}
    \hfill
    \includegraphics[width=0.32\textwidth]{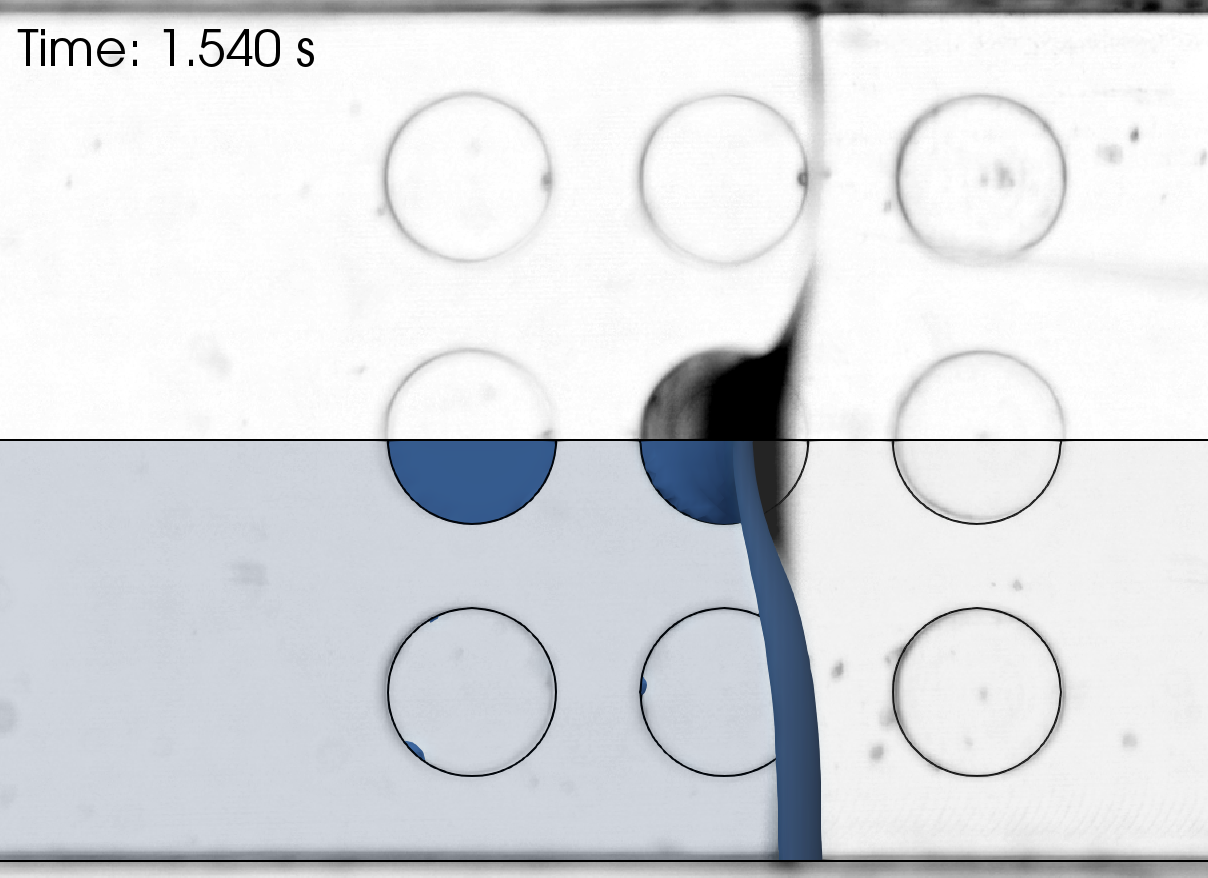}
    \includegraphics[width=\textwidth]{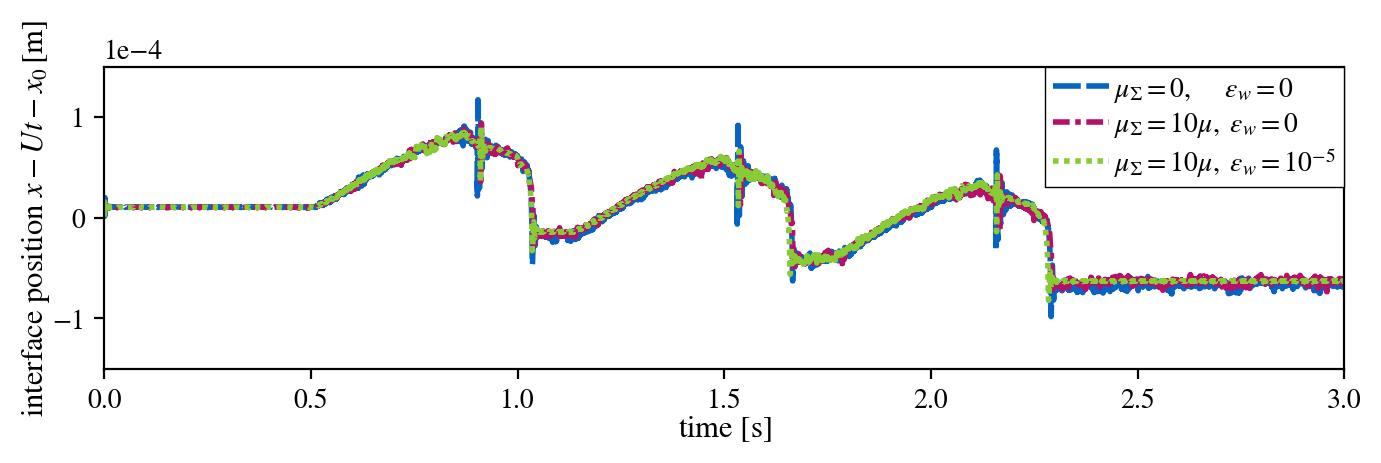}
    \hfill
    \includegraphics[width=\textwidth]{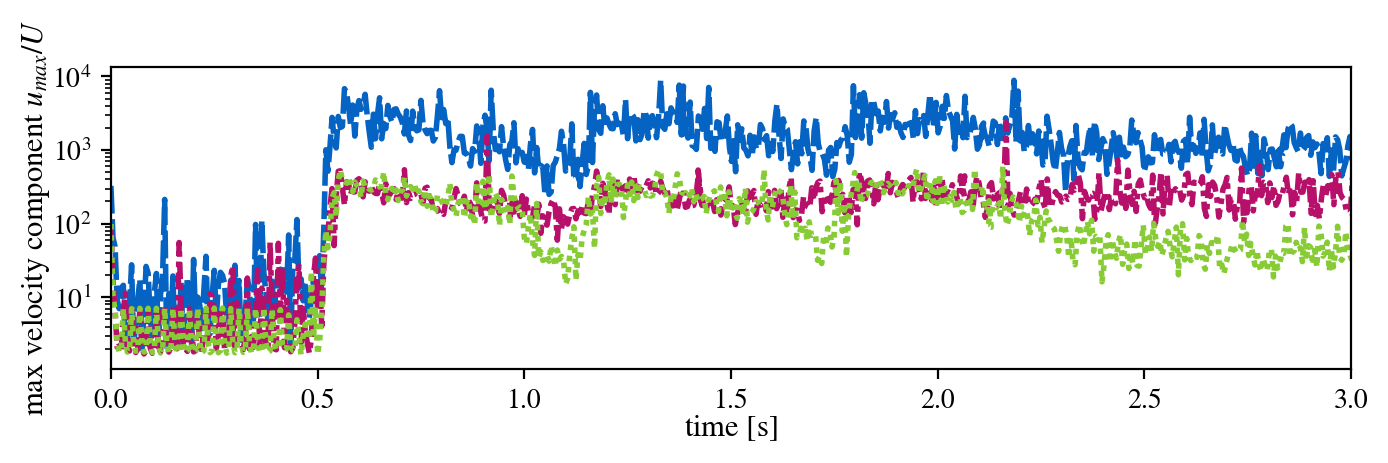}
    \hfill
    \caption{Results for 3x3 cavity flow with $U=\SI{0.001}{m/s}$ and contact angle $\theta=\SI{90}{\degree}$ and $\theta=\SI{120}{\degree}$ at the bottom and top boundary, respectively. Top: qualitative comparison. Wisp removal active.}
    \label{fig:06_results_filling}
\end{figure}
\subsubsection{Traversing interface with cavity filling}
\label{subsubsec:06_filling}
For the second test case, the inlet velocity is chosen as $U=\SI{0.001}{m/s}$ and the corresponding contact angle at the wall is $\theta=\SI{90}{\degree}$.
As cover plate, a polycarbonate plate with increased roughness is used instead of PMMA to improve the filling behavior at the cavities. 
Therefore, at the upper wall, the contact angle is $\theta=\SI{120}{\degree}$.
The visual comparison of the experiment and the simulation with $\mu_\Sigma=0$ is given in the upper subfigure of~\cref{fig:06_results_filling} for three points in time.
Here, due to the reduced velocity and corresponding lower apparent contact angle as well as the high contact angle at the upper wall, the cavities are filled with liquid in the experiments.
The contact line is pinned at the cavity edges first before snapping through the cavities at high speed, leaving small air bubbles behind.
In the simulations, this behavior is partly captured.
The cavities near the wall are filled except for small remaining air bubbles.
However, in the cavities at the channel center, an air cushion remains in the simulation, similar to the first test case, cf.~\cref{subsubsec:06_notfilling}.
Since the cavity filling is a highly dynamic process, it can be assumed that the constant contact angle employed in the simulations is not suited to represent the dynamic contact angle, which varies throughout the filling process.
The dynamics of the contact line movement are also visible in the center figure of \cref{fig:06_results_filling}, where the temporal evolution of the interface position is depicted.
In general, the progression of the interface is similar to the one seen in~\cref{fig:06_results_notfilling}, with exception of a high oscillation when a cavity is filled, e.g. at $t=\SI{0.9}{s}$.
Here, due to the high contact line velocity, the CFL time step criterion is temporarily breached, which can be assumed to be the cause for the interface oscillations.\notes{todo - explain more?}
With the artificial viscosity model applied, a decrease of the oscillation magnitude is achieved.
\par 
The effect of the artificial viscosity model and wisp removal algorithm on spurious currents is displayed in the bottom subfigure of~\cref{fig:06_results_filling}.
The maximum velocity component in the domain is reduced by the artificial viscosity, parallel to the results discussed in~\cref{subsubsec:06_notfilling}.
The reduction of spurious currents in the first stage of the flow due to the wisp removal also matches the previously mentioned results.
A further positive effect of the wisp removal on the spurious currents is observed during the cavity filling process.
Wisps occurring during the filling are filtered by the wisp removal algorithm, eliminating a source of spurious currents in the domain.\notes{explain better.}
The mass loss due to the wisp removal remains below~$m < \SI{0.005}{} V \rho\_{air}$, with V being the domain volume.
\notes{
\begin{itemize}
    \item reproducability tests - not alyways completely filled, size of air bubbles varied.
    \item 
\end{itemize}
}
%
\notes{
	\begin{itemize}
        \item in the filling stage before the cavities are reached, peaks in the maximum velocity appear
        \item they have no significant influence on the interface movement and are localized in a single cell.
        \item their location is in the interface viscinity at the symmetry plane.
        \item They only appear with active wisp correction.
        \item An improved wisp correction algorithm might employ a similar method to the one shown here, but using the corrected alpha field only for advection and keeping the original uncorrected alpha field in the momentum equation.
        \item This could impede drastic jumps in the velocity due to removed wisps.
        \item Further investigations suggest that an increased wisp tolerance increses the appearence of velocity jumps.
        \item A more thorough investigation is beyond the scope of this work and will be pursued in the future.
        \item (should we ignore wisps at the symmetry boundary?)
	\end{itemize}
}

\section{Conclusions}
\label{sec:07_conclusions}
%
To investigate flow regimes and species transport in microsctructures present in LoC applications, a reduction of spurious currents in separated two-phase flows is crucial.
To this end, the influence of an artificial viscosity model in the simulation of the interface traversing through microstructures was investigated in this work.
%
The artificial viscosity model presented by Denner et al.~\cite{denner_artificial_2017} was implemented as an OpenFOAM fvOption and validated against the results from~\cite{denner_artificial_2017} for a 2D oscillating capillary wave.
Further hydrodynamic and wetting validations with fluid pairings and flow regimes relevant for LoC applications were considered.
Overall, the artificial viscosity showed good performance in dampening spurious currents in the vicinity of the interface.
The thorough validation showed that it can be beneficial to increase the artificial viscosity value even beyond the Raessi model given in~\cite{raessi_semi-implicit_2009}.
In the regime relevant for LoC applications, it can be chosen up to two orders of magnitude higher than the larger physical viscosity in the domain.
However, the appropriate range of artificial viscosity depends on the boundary conditions and the fluid properties.
In dynamic surface-tension driven cases, high values of the artificial viscosity can reduce the solution quality with excessive dissipation.
The standard artificial viscosity formulation given by Raessi et al.~\cite{raessi_semi-implicit_2009} does not account for the sensitivity to fluid properties and boundary conditions.
Therefore, a general model for the artificial viscosity that extends the Raessi model by considering the physical viscosities would be beneficial to make the model operational for different flow regimes without requiring calibration pre-tests for the selection of a constant viscosity value.
Another promising approach is the data-driven modeling of a case-specific optimal artificial viscosity by employing machine learning.
\par In the comparison of the solvers, interFlow-fPa was shown to be overall superior to interIsoFoam in the considered cases regarding spurious currents.
\par 3D simulations of a fluid interface traversing through a 3x3 microcavity array were able to predict the temporal evolution of the interface to a satisfactory extent.
It was shown that the artificial viscosity can reduce the magnitude of spurious currents by over one order of magnitude without significantly influencing the interface progression.
Deviations from the experimental results might be accounted for by introducing a dynamic contact angle model in the simulations.
This will be investigated in future work.
Finally, the wisp removal algorithm was shown to be a method to reduce spurious currents in simulations with complex geometry in the IsoAdvector-plicRDF geometrical VoF method~\cite{roenby_computational_2016}.
The presented results will serve as a basis for further quantitative investigations of 3D transient microcavity flow regimes in the context of LoC applications.
%
\section{Acknowledgments}
The last author acknowledges the funding by the German Research Foundation (DFG): 
July 1 2020 - 30 June 2024
funded by the German Research Foundation (DFG) - Project-ID 265191195 - SFB 1194.

\appendix
\section{Appendix}
\label{sec:sample:appendix}
\subsection{2D capillary wave results for interIsoFoam}
\label{app:denner_interIso}
\begin{figure}[!h]
    \includegraphics[width=0.44\textwidth]{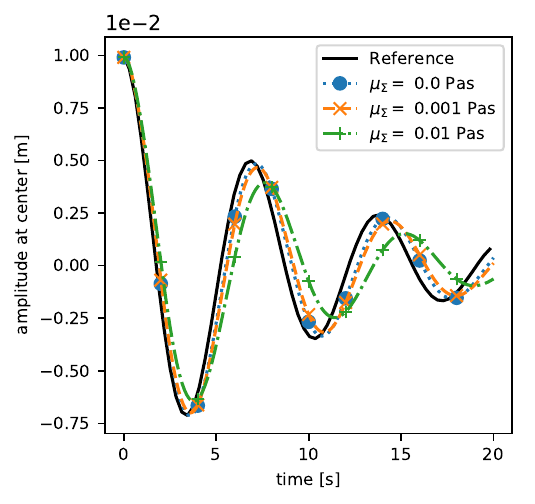}
    \caption{The temporal evolution of the oscillation amplitude of the capillary wave with $\lambda = \SI{1}{m}$ with different values of interface viscosity and solver interIsoFoam.}
    \label{fig:app_1}
\end{figure}
\begin{figure}[!h]
    \hspace{0.2cm}
    \includegraphics[width=0.44\textwidth]{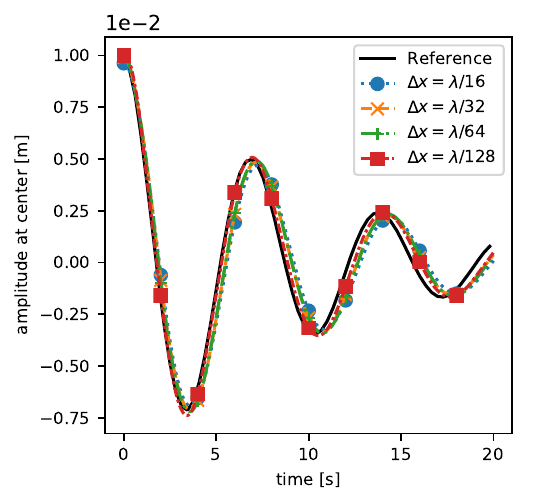}
    \hfill
    \includegraphics[width=0.44\textwidth]{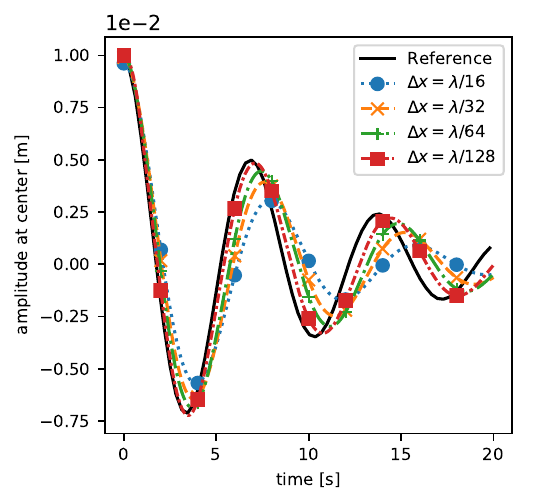}
    \hspace{0.2cm}
    \caption{Results obtained with interIsoFoam for mesh study of 2D capillary wave oscillation amplitude. Left: results with interface viscosity $\mu_\Sigma = \SI{0.}{}$ for different mesh resolutions, denoted by the number of cells per wavelength. Right: mesh study with interface viscosity $\mu_\Sigma = \SI{0.01}{Pa s}$. }
    \label{fig:app_2}
\end{figure}
In Figures~\ref{fig:app_1} and~\ref{fig:app_2} the results of the 2D capillary wave setup described in~\cref{subsubsec:validation-hydrodynamic-wave} are given for the solver interIsoFoam. 
The influence of the artificial viscosity on the oscillation amplitude matches the descriptions given in~\cref{subsubsec:validation-hydrodynamic-wave}.
 \bibliographystyle{elsarticle-num} 
 \bibliography{bib1}





\end{document}